\documentclass[]{emulateapj}
\usepackage{natbib}
\usepackage{apjfonts}
\usepackage{epsfig}

\newcommand{\Mpc}{\mathrm{Mpc}}
\newcommand{\Msun}{M_{\sun}}
\newcommand{\Omegam}{\Omega_{\mathrm{m}}}
\newcommand{\Omegab}{\Omega_{\mathrm{b}}}
\newcommand{\Omegar}{\Omega_{\mathrm{r}}}
\newcommand{\Omegal}{\Omega_{\Lambda}}
\def\LCDM {{\ifmmode\Lambda\mathrm{CDM}\else$\Lambda\mathrm{CDM}$\fi} }

\def\deg      {{\ifmmode~\mathrm{deg}\else$~\mathrm{deg}$\fi} } %%% Overwrites TeX \deg
\def\mag      {{\ifmmode~\mathrm{mag}\else$~\mathrm{mag}$\fi} } %%% Overwrites TeX \mag

\newcommand{\pa}{p_{\mu}}
\newcommand{\pb}{p_{\nu}}
\newcommand{\Fab}{F_{\mu\nu}}
\newcommand{\Cab}{C_{\mu\nu}}

\newcommand{\Dij}{D_{ij}}
\newcommand{\Sij}{S_{ij}}
\newcommand{\Pij}{P_{ij}}
\newcommand{\bF}{\mathbf{F}}
\newcommand{\bC}{\mathbf{C}}
\newcommand{\bCp}{\mathbf{C}_{\mathrm{prior}}}
\newcommand{\bCc}{\mathbf{C}_{\mathrm{combo}}}
\newcommand{\bD}{\mathbf{D}}
\newcommand{\bP}{\mathbf{P}}
\newcommand{\bS}{\mathbf{S}}
\newcommand{\bk}{\mathbf{k}}
\newcommand{\bp}{\mathbf{p}}
\newcommand{\bx}{\mathbf{x}}
\newcommand{\br}{\mathbf{r}}
\newcommand{\Tr}{\mathrm{Tr}}
\newcommand{\pd}{\partial}
\newcommand{\dd}{\mathrm{d}}
\newcommand{\dr}{\delta r}
\newcommand{\sigmaDM}{\sigma_{\mathrm{DM}}}
\newcommand{\sigmaCV}{\sigma_{\mathrm{CV}}}
\newcommand{\Lya}{\mathrm{Ly}\alpha}
\newcommand{\sinc}{\mathrm{sinc}}
\newcommand{\logten}{\log_{10}}
\newcommand{\oangle}{\Theta}
\newcommand{\boangle}{\mathbf{\Theta}}

\newcommand{\UDFGO}{UDF~GO}
\newcommand{\GOODSERS}{GOODS~ERS}
\newcommand{\UDFDeep}{Survey~A}
\newcommand{\GOODSDeep}{Survey~B1}
\newcommand{\GOODSMedium}{Survey~B2}
\newcommand{\GOODSShallow}{Survey~B3}
\newcommand{\GOODSAll}{Surveys~B1,~B2,~and~B3}
\newcommand{\AllSurveys}{Surveys~A,~B1,~B2,~B3,~C,~and~D}
\newcommand{\SEDS}{Survey~C}

\newcommand{\COSMOS}{Survey~D}

\newcommand{\fcomp}{f_{\mathrm{comp}}}
\newcommand{\focc}{f_{\mathrm{occ}}}
\newcommand{\zmin}{z_{\mathrm{min}}}
\newcommand{\zmax}{z_{\mathrm{max}}}
\newcommand{\Mmax}{M_{\mathrm{max}}}
\newcommand{\Atot}{A_{\mathrm{tot}}}
\newcommand{\bave}{\langle b\rangle}
\newcommand{\Nfields}{N_{\mathrm{fields}}}
\newcommand{\Norbits}{N_{\mathrm{orbits}}}
\newcommand{\HAB}{H_{\mathrm{AB}}}
\newcommand{\FSIXTY}{F160W}
\newcommand{\FFOURTY}{F140W}

\newcommand{\rhobar}{\bar{\rho}}
\newcommand{\nbar}{\bar{n}}
\newcommand{\nh}{n_{\mathrm{h}}}
\newcommand{\nhi}{n_{\mathrm{h},i}}
\newcommand{\phistar}{\phi_{\star}}
\newcommand{\Mstar}{M_{\star}}
\newcommand{\MUV}{M_{\mathrm{UV}}}
\newcommand{\deltac}{\delta_{c}}

\newcommand{\What}{\hat{W}}
\newcommand{\Whats}{\hat{W}^{\star}}

\shorttitle{}
\shortauthors{}

\begin{document}

\title{Estimating Luminosity Function Constraints from High-Redshift Galaxy Surveys}
\author{Brant E. Robertson\altaffilmark{1}}

\altaffiltext{1}{Hubble Fellow, brant@astro.caltech.edu}
\affil{Astronomy Department, California Institute of Technology, 
MC 249-17, 1200 East California Boulevard, Pasadena, CA 91125}

\begin{abstract}
The installation of the Wide Field Camera 3 (WFC3) on the Hubble Space Telescope (HST)
will revolutionize the study of high-redshift galaxy populations.  
Initial observations of the HST Ultra Deep Field (UDF) have yielded multiple
$z\gtrsim7$ dropout candidates.  Supplemented by the Great Observatories Origins Deep
Survey (GOODS) Early Release Science (ERS) and further UDF pointings, these data will
provide crucial information about the most distant known galaxies.
However, achieving tight constraints on the $z\sim7$ galaxy luminosity function (LF) will require
even more ambitious photometric surveys.
Using a Fisher matrix approach to
fully account for Poisson and cosmic sample variance, as well as covariances in the
data, we estimate the uncertainties on LF parameters achieved by
surveys of a given area and depth.  Applying this method to WFC3 $z\sim7$
dropout galaxy samples, we forecast the LF parameter uncertainties for a variety of
model surveys.  We demonstrate that performing a wide area ($\sim 1\deg^{2}$) survey to $\HAB\sim27$ depth 
or increasing the UDF depth to $\HAB\sim30$ 
provides excellent constraints on the high-$z$ LF when combined with the existing \UDFGO~ and \GOODSERS~data.
We also show that the shape of the matter power spectrum may limit the possible 
gain of splitting wide area ($\gtrsim0.5\deg^{2}$) high-redshift surveys into multiple fields to probe 
statistically independent regions; the increased root-mean-squared density fluctuations
in smaller volumes mostly offset the improved variance gained from independent samples.
\end{abstract}


%%%%%%%%%%%%%%%%%%%%%%%%%%%%%%%%%%%%%%
%
%	Introduction
%
%%%%%%%%%%%%%%%%%%%%%%%%%%%%%%%%%%%%%%
\section{Introduction}
\label{section:introduction}

Recent studies using Hubble Space Telescope (HST) Wide Field
Camera 3 (WFC3) observations 
have discovered tens of candidate galaxies at redshifts 
$z\gtrsim7$ 
\citep{bouwens2009a,bouwens2010b,bouwens2010a,oesch2010a,oesch2010b,bunker2009a,mclure2009a,yan2009a,wilkins2009a,labbe2009a,labbe2010a,finkelstein2009a}.
The new WFC3 observations have broadened our knowledge of 
the highest redshift galaxies yet found, complementing $z\gtrsim7$
galaxy searches 
with the
Near Infrared Camera and Multi-Object Spectrometer 
\citep{kneib2004a,bouwens2004a,bouwens2005a,egami2005a,henry2007a,henry2008a,henry2009a,richard2008a,bradley2008a,bouwens2008a,zheng2009a,oesch2009a,gonzalez2009a},
ground-based dropout selections \citep{richard2006a,stanway2008a,ouchi2009b,hickey2009a,castellano2009a},
and narrow band Lyman-$\alpha$ ($\Lya$) emission surveys
\citep{parkes1994a,kodaira2003a,santos2004b,willis2005a,willis2008a,taniguchi2005a,stark2006a,iye2006a,kashikawa2006a,stark2007a,cuby2007a,ota2008a,ouchi2009a,hibon2009a,sobral2009a}.
The existence of star-forming galaxies at $z\gtrsim7$ has been
well-established by these studies, and the importance of
these high-redshift galaxies for reionization 
and subsequent galaxy formation at lower redshifts will
likely motivate the dedication of 
large telescope allocations to
detailing their abundance.  The purpose of this paper
is to develop a method to rapidly compare 
possible photometric survey strategies for detecting
large numbers of $z\sim7$ galaxies,
and forecast constraints on the $z\sim7$ galaxy
luminosity function (LF) achieved by different HST survey designs.

A determination of the constraining power a survey can
obtain requires an estimate of the uncertainty in the
abundance of galaxies as a function of luminosity.
In addition to the Poisson uncertainty inherent in 
galaxy counts, cosmic sample variance induced by
density fluctuations and galaxy clustering must
be accounted for
\cite[see, e.g.,][]{newman2002a,somerville2004a,stark2007b}.
A particularly powerful approach for estimating 
these uncertainties and determining the resulting
potential constraints
on the luminosity function of galaxies in the UDF
was presented by \cite{trenti2008a}. 
These authors used
cosmological simulations to determine
the abundance and spatial distribution of 
dark matter halos and then applied a model for
the halo mass-to-light ratio to determine the
abundance of galaxies of a given luminosity.
Poisson and 
sample variance uncertainties were
estimated by drawing pencil beam realizations of the
survey from the cosmological volume \citep[see also][]{kitzbichler2007a}. 
A maximum-likelihood approach was then used
to study constraints on the luminosity function
for various dropout selections.

The maximum-likelihood estimation of LF parameters based on 
mock catalogues can account for detailed selection effects 
and spatial correlations in addition to the Poisson and
sample variance uncertainties.  Such simulations have 
clear advantages
for estimating the completeness of magnitude-
limited surveys or understanding systematic effects introduced
by dropout color selections.  However, the need for
halo catalogues from cosmological simulations introduces two
limitations.  First, the volume of the simulation should probe
many independent realizations of the modeled survey. 
However, future high-redshift galaxy surveys with WFC3 or other 
instruments may take the form of deeper
versions of surveys like 
the Spitzer Extended Deep Survey \citep[SEDS;][]{fazio2008a} Exploration Science
program
or the Cosmic Evolution Survey \citep[COSMOS;][]{scoville2007a,scoville2007b}.
These surveys are each $\sim1\deg^{2}$ or larger,
and large volume 
($L\gtrsim200h^{-1}\Mpc$) cosmological simulations are required
to probe multiple independent samples of the surveys' high-redshift
galaxy populations.  
For instance, a $\sim1\deg^{2}$ survey at $6.5\lesssim z\lesssim 7.5$
has a comoving volume of $V\approx3\times10^{6} h^{-3} \Mpc^{3}$.
The largest simulation used by \cite{trenti2008a} would provide
less than two independent samples of such a volume, and even the 
Millennium Simulation \citep{springel2005a} with a comoving box size 
$L=500h^{-1}\Mpc$ would only provide $\sim40$ independent samples
of such a wide high-redshift survey.  Second, the method requires access to and
manipulation of cosmological simulation results.  This requirement
may pose an unwanted computational overhead for those interested in rapid estimates
and comparisons of potential constraints from a wide range of 
survey designs.

A simpler methodology
for estimating survey constraints on the abundance of high-redshift
galaxies that does not directly require halo catalogues from
cosmological simulations is
therefore desirable for performing rapid comparisons of survey designs.
Hence, we seek an approximate method for forecasting LF parameter
constraints that relies on 
descriptions of galaxy
and dark matter halo abundance and clustering, Poisson and
cosmic sample variance, and parameter covariances that are
analytical or easily calculable through numerical methods.
We utilize a simple model for describing the
clustering of $z\sim7$ galaxies based on
fiducial empirical estimates of the high-redshift galaxy
luminosity function \citep{oesch2010a} and abundance
matching between galaxies and dark matter halos \citep[e.g.,][]{conroy2006a,conroy2009a}.
We then adopt a common approach to
translate galaxy clustering and matter fluctuations into
an estimate of the cosmic sample variance \citep[see the various calculations in, e.g.,][]{newman2002a,somerville2004a,stark2007b,trenti2008a}.
With this estimate of the sample variance and Poisson uncertainty
from an assumed fiducial model for the abundance of galaxies,
we use a Fisher matrix formalism to characterize the likelihood 
function and estimate $z\sim7$ luminosity function parameter constraints. 
The presented methodology is fast and flexible, and
can be used with appropriate extensions,  
to estimate constraints on galaxy abundance for
other survey sample selections and redshifts.

Motivated by the exciting recent HST WFC3 results,
we focus on modeling $z\sim7$ dropout survey
designs.
While we choose to study broadband searches
for high-redshift galaxies, narrow band surveys
for high-redshift $\Lya$ emission present
another interesting class of survey designs.
The rapid progress in detecting increasing
numbers of high-redshift $\Lya$ emitters 
using narrow band surveys has 
motivated
theoretical efforts both to understand 
and predict the abundance of $\Lya$ emitters.
The observable properties of the
high-redshift $\Lya$ emitter population 
are particularly difficult to model owing
to the uncertain escape fraction, 
intergalactic medium absorption and
other radiative transfer effects, as well as 
uncertainties in our knowledge of the
galaxy formation process
\citep[e.g.,][]{haiman2002a,santos2004a,barton2004a,wyithe2005a,le_delliou2006a,hansen2006a,dave2006a,furlanetto2006a,tasitsiomi2006a,mcquinn2007a,nilsson2007a,mao2007a,kobayashi2007a,kobayashi2010a,stark2007b,mesinger2008a,fernandez2008a,tilvi2009a,dayal2009a,dayal2010a,samui2009a}.
While the astrophysics involved in these
studies are tremendously interesting and
provide another route to probe high-redshift
galaxies, we will only examine the
statistical constraining power of various
broadband survey designs and will not
attempt to model the $\Lya$ emitter population.

This paper is organized as follows.
Forecasting constraints on $z\sim7$ LF function parameters requires
a model of the sources of error and covariances in
the data.  In \S \ref{section:variances}, we discuss
sample variance uncertainties owing cosmic density fluctuations.
In \S \ref{section:fisher_matrix}, we review the Fisher
matrix formalism and show how to apply the formalism to 
forecast LF parameter uncertainties accounting for 
cosmic sample and Poisson variances.  To perform actual
forecasts for the $z\sim7$ LF, we review existing HST 
WFC3 survey data in \S \ref{section:existing_surveys} and
define fiducial model surveys in \S \ref{section:model_surveys}.
In \S \ref{section:results}, we combine the expected
constraints from existing surveys with forecasts
of luminosity function constraints from model surveys.
We discuss our results
and possible caveats in \S \ref{section:discussion}, and
summarize and conclude
in \S \ref{section:summary}.

Throughout, we work in the context of a \LCDM cosmology
consistent with joint constraints from the 
5-year Wilkinson Microwave Anisotropy Probe,
Type Ia supernovae, Baryon Acoustic Oscillation, and Hubble
Key Project data 
\citep{freedman2001a,percival2007a,kowalski2008a}. Specifically,
we adopt a Hubble parameter $h=0.705$, matter density
$\Omegam=0.274$, dark energy density $\Omegal=0.726$, baryon density
$\Omegab=0.0456$, relativistic species density $\Omegar=4.15\times10^{-5}$,
 spectral index $n_{s}=0.96$, and root-mean-squared density fluctuations in 8~$h^{-1}$Mpc-radius 
spheres
of $\sigma_{8}=0.812$ \citep{komatsu2009a}.  We report all magnitudes in the AB system \citep{oke1983a}.

%%%%%%%%%%%%%%%%%%%%%%%%%%%%%%%%%%%%%%
%
%	Cosmic Variance and the Power Spectrum
%
%%%%%%%%%%%%%%%%%%%%%%%%%%%%%%%%%%%%%%
\section{Poisson Uncertainty, Cosmic Sample Variance, and the \LCDM Power Spectrum}
\label{section:variances}

We wish to evaluate the relative merits of
various galaxy survey designs in terms of their
ability to constrain the galaxy luminosity
function.  To perform this evaluation, we
must determine the quality of each 
design in terms of the number of galaxies
of a given luminosity the survey will discover
(the Poisson variance) and the
intrinsic scatter expected for the survey
volume given variations in the cosmological 
density field (the cosmic sample variance).
This section of the paper formally defines
each source of uncertainty and describes how
these variances are calculated.

We define cosmic sample variance as the fluctuations
in a volume-averaged quantity owing to
density inhomogeneities seeded by the matter power spectrum.
We will use the terms ``cosmic variance'' and ``sample variance'' interchangeably,
but elsewhere in the literature cosmic variance is taken to equal
the sample variance only in the limit of the entire volume of the universe
\citep[e.g.,][]{hu2003a}.

%%%%%%%%%%%%%%%%%%%%%%%%%%%%%%%%%%%%%%
%
%	Dark Matter Density Variance
%
%%%%%%%%%%%%%%%%%%%%%%%%%%%%%%%%%%%%%%
\subsection{Dark Matter Density Variance}
\label{subsection:sigmaDM}

Density fluctuations, or differences between the local matter density
$\rho_{m}(\bx)$ and the mean matter density $\rhobar_{m}$, can be described 
in terms of a local matter 
overdensity $\delta(\bx) \equiv [\rho_{m}(\bx) - \rhobar_{m}]/\rhobar_{m}$. 
Consider a survey of comoving volume $V$ at redshift $z$.
For ``unbiased'' quantities measured within $V$ that
spatially cluster like the dark matter,
such that the two-point correlation function $\xi(\br)$ is identical to the
dark matter correlation function $\xi_{m}(\br) = \langle \delta(\bx)\delta(\bx+\br)\rangle$,
the cosmic variance is simply the dark matter
variance
\begin{eqnarray}
\label{eqn:sigma_DM}
D^{2}(z)\sigmaDM^{2} \equiv \langle \delta^{2}(\bx) \rangle 
	= D^{2}(z) \int \frac{\dd^{3} k}{(2\pi)^{3}} P(k) |\What(\bk,V)|^{2}, 
\end{eqnarray}
\noindent
where $\What(\bk,V)$ is the Fourier transform of the survey volume 
window function $W(\bx)$ (whose geometry
may introduce a dependence on the direction of the wavenumber $\bk$), $D(z)$
is the linear growth function,  and
$P(k)$ is the isotropic linear $\LCDM$ power spectrum. 
To calculate $P(k)$, we use the transfer function of \citet{eisenstein1998a} 
that includes the effects of baryons.  We ignore possible nonlinear corrections to the
power spectrum \citep[e.g.,][]{peacock1996a,smith2003a}.
The window function is normalized such that $\int \dd^{3} x W(\bx) = 1$.
For a spherical volume of comoving radius $R=8h^{-1}\Mpc$, Equation \ref{eqn:sigma_DM} would
provide $\sigmaDM=\sigma_{8}$ at redshift $z=0$.
The linear growth function 
\begin{equation}
\label{eqn:growth_function}
D(z) = D_{0} H(z) \int_{z}^{\infty} \frac{(1+z') \dd z'}{H^{3}(z')}
\end{equation}
\noindent
has a normalization constant $D_{0}$ chosen such that $D(z=0) = 1$.
The Hubble parameter
\begin{eqnarray}
\label{eqn:hubble_parameter}
H(z) &=& H_{0}[\Omegar(1+z)^{4} + \Omegam(1+z)^{3} \nonumber \\
&&+ (1-\Omegam-\Omegal-\Omegar)(1+z)^{2} + \Omegal]^{1/2}
\end{eqnarray}
\noindent
describes the rate of change of the universal scale factor $H\equiv \dot{a}/a$ as
a function of the matter density $\Omegam$, relativistic species density $\Omegar$, and dark 
energy density $\Omegal$ (taken to be a cosmological constant).

%%%%%%%%%%%%%%%%%%%%%%%%%%%%%%%%%%%%%%
%
%	Dark Matter Halo Abundance
%	and Clustering
%
%%%%%%%%%%%%%%%%%%%%%%%%%%%%%%%%%%%%%%
\subsection{Dark Matter Halo Abundance and Clustering}
\label{subsection:halos}

For galaxy surveys, where quantities of interest depend on 
the abundance and clustering of galaxies,
the sample variance will depend on the bias of dark matter halos
hosting the observed systems.
We can define the bias in terms of the correlation function as $b^{2}  = \xi_{h}/\xi_{m}$
where $\xi_{h}$ is the correlation function of dark matter halos.
Given a halo mass function, the bias of halos with mass $m$ can be estimated using the
peak-background split formalism \citep[e.g.,][]{kaiser1984a,mo1996a,sheth1999a}
or measured directly from the simulations via the halo correlation function
or halo power spectrum.  We adopt the latter approach.

We use the dark matter halo mass function measured by \citet{tinker2008a} 
from a large suite of cosmological simulations 
\citep{kravtsov2004a,warren2006a,crocce2006a,gottloeber2007a,yepes2007a}.
The \citet{tinker2008a} mass function can be written as a function of the 
dark matter halo mass $m$ in terms of the
``peak height'', 
\begin{equation}
\label{eqn:peak_height}
\nu = \frac{\deltac}{ D(z) \sigma(m)},
\end{equation}
\noindent
 where $\deltac=1.686$ is the
spherical collapse barrier \citep[see, e.g,][]{gunn1972a,bond1996a},
$D(z)\sigma(m)$ is the square root of the dark matter variance (Equation \ref{eqn:sigma_DM})
evaluated in a spherical volume of comoving radius $R = (3m/4\pi\rhobar_{m})^{1/3}$.
Here, $\rhobar_{m}$ is the background matter density.  The linear growth
function $D(z)$ is given by Equation \ref{eqn:growth_function}.

With the definition of
peak height $\nu$ in Equation \ref{eqn:peak_height}, the \cite{tinker2008a} halo
mass function can be written
\begin{equation}
\label{eqn:mass_function}
\frac{\dd n_{\mathrm{h}}}{\dd m} = \frac{\rhobar_{m}}{m} f(\nu) \frac{\dd \nu}{\dd m},
\end{equation}
\noindent
where the function
\begin{equation}
\label{eqn:fcd}
f(\nu) = \alpha\left[ 1 + (\beta\nu)^{-2\phi}\right]\nu^{2\eta}e^{-\gamma\nu^{2}/2},
\end{equation}
\noindent
is often called the ``first crossing distribution.''  \citet{tinker2008a} find that
the parameter values
$\alpha = 0.245$, $\beta=0.757$, $\gamma=0.853$, $\phi=-0.659$, and $\eta=-0.341$
fit well the $z=2.5$ simulated mass function measured for halos
defined with a spherical overdensity $\Delta=200$ relative to the background matter density
\citep[see also \S 4 of ][]{tinker2010a}.  The abundance of
dark matter halos more massive than $m$ is then just $\nh(> m) = \int_{m}^{\infty} (\dd \nh/ \dd m) \dd m$.

We choose the \citet{tinker2008a} mass function because it is accurate
to $\lesssim 5\%$ for halos in the mass range $10^{11} h^{-1}\Msun\leq m \leq 10^{15}h^{-1}\Msun$
at redshift $z=0$, and improves on previous approximations by $10-20\%$ \citep[c.f.,][]{sheth1999a}.
\citet{tinker2008a} demonstrate that the halo mass function does not have a
redshift-independent, universal form, and that the normalization of the
first crossing distribution evolves at the $20-50\%$ level between $z=0$ and $z=2.5$.
However, the halo mass function has not been calibrated at the redshifts
of interest ($z\gg2.5$) and following the advice in \S 4 of \citet{tinker2008a}
we will use the $z=2.5$ first crossing distribution as the best available approximation.\footnote{
While we account for the redshift-dependent abundance of dark matter halos at
the mean redshift of the survey,
we ignore the evolution of the dark matter halo abundance over the
redshift interval of the survey volume. See, e.g., \citet[][]{munoz2008a} for
a quantification of the effect of this evolution on the variance of inferred halo number densities.}
We note that using any other previously published mass
function from the literature \citep[e.g.,][]{sheth1999a} will therefore introduce
an unknown error in the abundance of halos at high redshifts.
\citet{tinker2008a} estimate this error could be as large as $\sim20-50\%$ for galaxy-sized halos.\footnote{
We find that when using the \citet{sheth1999a} mass function 
and the corresponding \citet{sheth2001a} bias function,
the marginalized errors calculated in
sections \S \ref{section:existing_surveys} and \ref{section:results}
degrade by $\sim5\%$ relative to the results obtained with the
\citet{tinker2008a} mass function and \citet{tinker2010a} bias
function.  This difference quantifies how the results of the
presented method depend on the choice for the halo 
mass and bias functions.}

For the bias $b$ relating halo and dark matter clustering, we will
use the results of \citet{tinker2010a} who measure the halo bias
as a function of peak height $\nu$ in a manner consistent with the
halo mass function of \citet{tinker2008a}.  The bias function $b(\nu)$
is constrained by the halo first crossing distribution $f(\nu)$ by requiring
that dark matter is not biased against itself, i.e.,
\begin{equation}
\label{eqn:bias_constraint}
\int b(\nu) f(\nu) \dd \nu = 1.
\end{equation}
\noindent
Under this constraint, \cite{tinker2010a} find that the
fitting function
\begin{equation}
\label{eqn:bias}
b(\nu) = 1 - A\frac{\nu^{a}}{\nu^{a} + \deltac^{a}} + B\nu^{b} + C\nu^{c}
\end{equation}
\noindent
with parameters $A=1.0$, $a=0.1325$, $B=0.183$, $b=1.5$, $C=0.265$, and $c=2.4$
provides an accurate match to the bias of dark matter halos defined with
a spherical overdensity of $\Delta=200$ relative to the background matter density.
As demonstrated by \citet{tinker2010a}, Equation \ref{eqn:bias} reproduces the
simulated halo clustering better than the analytical formulae 
of \citet{mo1996a} or \citet{sheth2001a} calculated using the peak-background
split formalism.
\citet{tinker2010a} find that the bias $b(\nu)$ as a function of peak height
$\nu$ is nearly redshift-independent, and we
will adopt Equation \ref{eqn:bias} for $b(\nu)$ at all redshifts.

%%%%%%%%%%%%%%%%%%%%%%%%%%%%%%%%%%%%%%
%
%	Galaxy Abundance and Clustering
%
%%%%%%%%%%%%%%%%%%%%%%%%%%%%%%%%%%%%%%
\subsection{Galaxy Abundance and Clustering}
\label{subsection:galaxies}

Our main premise is to use the Fisher matrix 
approach to estimate the constraints on a
model for the abundance of galaxies that 
reproduces well the observed source counts.
The definition of the model for galaxy
abundance is therefore important.  Further,
the clustering of galaxies directly influences
the covariances of the data and knowledge of
the galaxy spatial distribution is therefore
also quite important.  Since we are interested
in the characteristics of the high-redshift
galaxy population for which little clustering
information is known, we will estimate the
galaxy clustering bias by associating 
luminous galaxies with dark matter halos of
similar comoving abundance and assigning those
galaxies the bias of their associated halos
(as provided by Equation \ref{eqn:bias}).  
When more detailed clustering information is
available, as is the case at lower redshifts,
additional constraints on the
connection between galaxy and halo populations
are attainable \citep[see, e.g.,][]{lee2009a}.

We will adopt the commonly-used
\citet{schechter1976a} model for the abundance
of galaxies.  Specifically,
the expected number density $\nbar_{i}$ of galaxies in the $i$-th luminosity bin
of width $\Delta M$ about magnitude $M_{i}$ can be written
\begin{equation}
\label{eqn:luminosity_function}
\nbar_{i} = \int_{M_{i}-\Delta M/2}^{M_{i}+\Delta M/2} \Phi(M)\dd M,
\end{equation}
\noindent
where the \citet{schechter1976a} function
\begin{eqnarray}
\Phi(M) = \frac{2}{5} \ln(10) \phistar \left[10^{\frac{2}{5}(\Mstar - M)}\right]^{\alpha+1} 
	\exp\left[ - 10^{\frac{2}{5}(\Mstar-M)}\right],
\end{eqnarray}
\noindent
describes the distribution of galaxy luminosities,
$\phistar$ is the
luminosity function normalization in comoving $\Mpc^{-3} \mag^{-1}$, $\Mstar$
is the characteristic galaxy luminosity in AB magnitudes, and $\alpha$ is the
faint-end slope.
We will often refer to the parameters of 
this \citet{schechter1976a} model in terms of the vector 
$\bp = [\logten\phistar,\Mstar,\alpha]$, and it is these parameters
for which we will forecast constraints.  The fiducial values for
the parameters $\bp$ used in the Fisher matrix calculation will be 
selected in \S \ref{section:existing_surveys}.

In a manner similar to Equation \ref{eqn:luminosity_function}, we
can also define the comoving abundance $\nbar_{L}$ of galaxies more
luminous than magnitude $M$ as $\nbar_{L}(<M) = \int_{-\infty}^{M} \Phi(M)\dd M$,
where the negative lower limit follows from the definition of magnitudes.

With a model for the abundance of galaxies, we will
associate galaxies with dark matter halos of similar
abundance to estimate the galaxies' spatial clustering.
The abundance $\nbar_{i}$ of galaxies in the range $M_{i} \pm \Delta M/2$ 
can be written as 
$\nbar_{i} \equiv \nbar_{L}(<M_{i} + \Delta M/2) - \nbar_{L}(<M_{i} - \Delta M/2)$.
We match the abundance of galaxies and halos as
\begin{equation}
\label{eqn:abundance_matching}
\nh(> m) \simeq \nbar_{L}(<M)
\end{equation}
\noindent
at the minimum and maximum luminosity of galaxies in each magnitude bin (e.g., 
$M = M_{i} + \Delta M/2$ and $M= M_{i} - \Delta M/2$), which provides
a mass range $m_{i} \pm \Delta m/2$ of halos with a similar abundance \citep[e.g.,][]{conroy2006a,conroy2009a}.  
The comoving
number density of these halos is simply
$\nhi\equiv \nh(>m_{i}-\Delta m/2)-\nh(>m_{i}+\Delta m/2)$, 
with $\nhi = \nbar_{i}$.  The resulting connection between 
galaxy luminosity and halo mass is simplistic, but more sophisticated
stellar mass-halo mass relations could be incorporated into our
approach when warranted by the constraining power of the available
data \citep[see, e.g.,][]{behroozi2010a}.  We adopt $\Delta M = 0.25$
mag throughout, but we have checked that our conclusions also hold for
$\Delta M=0.5$ or $\Delta M=0.1$.

The bias $b_{i}$ of galaxies in the range $M_{i}\pm\Delta M/2$ is then
approximated as the number-weighted average clustering of 
halos of mass $m_{i} \pm \Delta m /2$.
We can express $b_{i}$ as
\begin{eqnarray}
\label{eqn:bias_from_abundance}
b_{i} &=& \left[\int_{m_{i} - \Delta m/2}^{m_{i} + \Delta m/2} b(m) \frac{\dd \nh}{\dd m} \dd m\right]
\times \left[\int_{m_{i} - \Delta m/2}^{m_{i} + \Delta m/2} \frac{\dd \nh}{\dd m} \dd m\right]^{-1},
\end{eqnarray}
\noindent
where $b(m)\equiv b[\nu(m)]$ as defined in Equation \ref{eqn:bias}.

%%%%%%%%%%%%%%%%%%%%%%%%%%%%%%%%%%%%%%
%
%	Sample Covariance from Galaxy Abundance and Clustering
%
%%%%%%%%%%%%%%%%%%%%%%%%%%%%%%%%%%%%%%
\subsection{Sample Covariance from Galaxy Abundance and Clustering}
\label{subsection:sample_covariance}

Equation \ref{eqn:luminosity_function} provides the average expected 
abundance of galaxies in a survey, given our fiducial luminosity
function model $\Phi(M)$.  Owing to spatial density fluctuations on
large scales, the
actual measured number density of galaxies in the magnitude range $M\pm \Delta M/2$
at location $\bx$ will be 
\begin{equation}
\label{eqn:measured_number_density}
n_{i}(\bx,z) = \nbar_{i}[1 + b_{i} \delta(\bx,z)]
\end{equation}
\noindent
where $\delta(\bx,z)$ is the local linear overdensity and
the bias $b_{i}$ is determined in Equation \ref{eqn:bias_from_abundance}.
The large scale structure of the matter density field will
cause the galaxy counts to covary.  The sample covariance $\Sij$
between galaxies in the $i$-th and $j$-th magnitude bins is simply 
the average squared difference between the measured
galaxy density $n$ and the expected average galaxy density $\nbar$
for each bin.  We can then write the sample covariance as
\begin{equation}
\label{eqn:sample_covariance_matrix}
\Sij \equiv \langle (n_{i} - \nbar_{i})(n_{j} - \nbar_{j})\rangle,.
\end{equation}
\noindent
where the average is take over all $\Nfields$ fields of the survey.
Given Equations \ref{eqn:sigma_DM}, \ref{eqn:bias}, \ref{eqn:luminosity_function}, \ref{eqn:bias_from_abundance}, and \ref{eqn:measured_number_density},
we can evaluate the elements of the sample covariance matrix $\bS$ as
\begin{equation}
\label{eqn:sample_covariance}
\Sij  = \frac{b_{i} b_{j} \nbar_{i} \nbar_{j}}{\Nfields} D^{2}(z)\int \frac{\dd^{3} k}{(2\pi)^{3}} \What_{i}(\bk) \Whats_{j}(\bk) P(k)
\end{equation}
\noindent
where $\What_{i}(\bk)$ is the $k$-space window function for the survey field volume of the $i$-th 
magnitude bin.  Depending on, e.g., the redshift distribution of sources with different magnitudes,
or some luminosity-dependent completeness, we could have $\What_{i}(\bk)\ne\What_{j}(\bk)$ in general.
However, throughout the rest of the paper we will consider only galaxy densities and variances 
within the entire effective survey volume, such that the elements of the sample covariance matrix 
$\bS$ refer to luminosity bins within the same volume of each field.  We will therefore write
$\What_{i}(\bk)\Whats_{j}(\bk) = |\What_{V}(\bk)|^{2}$, where 
\begin{equation}
\label{eqn:window_function}
\What_{V}(\bk) = \sinc\left(\frac{k_{x}r\Theta_{x}}{2}\right)\sinc\left(\frac{k_{y}r\Theta_{y}}{2}\right)\sinc\left(\frac{k_{z} \dr}{2}\right)
\end{equation}
is an approximate $k$-space window function for the effective volume $V$ of a survey at comoving radial distance
$r$, comoving radial width $\dr$, and rectangular area $\Theta_{x}\times\Theta_{y}$ in square radians.\footnote{A rectangular survey will have a larger
on-sky footprint at $r+\dr$ than at $r$.  For $\Theta_{x}$ and $\Theta_{y}$ of interest to this paper ($\lesssim1\deg$) and $\dr/r\ll1$, we have checked that 
both $W(\bx)$ and $\What(\bk)$ are well-approximated by the window function in Equation \ref{eqn:window_function} and its transform.}  The
function $\sinc(x/2)=2\sin(x/2)/x$ is the Fourier transform of the Heaviside $\Pi(x)$ unit box.   Similar window functions
were adopted by \cite{newman2002a} and \cite{stark2007b} in their estimates of cosmic variance. 

Unless otherwise specified, when discussing the cosmic sample 
variance uncertainty or error
we will refer to the averaged quantity 
\begin{equation}
\label{eqn:cosmic_variance}
\sigmaCV \equiv \bave D(z)\sigmaDM/\sqrt{\Nfields},
\end{equation}
\noindent
where $\bave$ is the average bias of all galaxies in a survey
field.  This quantity $\sigmaCV$ is the cosmic variance
uncertainty that is often reported for surveys,
but is distinct from the elements sample covariance matrix $\Sij$
since the latter involves the bias of galaxies in individual 
luminosity bins.

%%%%%%%%%%%%%%%%%%%%%%%%%%%%%%%%%%%%%%
%
%	Multiple Fields and Sample Covariance
%
%%%%%%%%%%%%%%%%%%%%%%%%%%%%%%%%%%%%%%
\subsubsection{Multiple Fields and Sample Covariance}
\label{subsubsection:multiple_fields}

Splitting a survey into $\Nfields$ multiple fields 
can reduce
the sample covariance in the combined data by probing
statistically independent regions in space.  The
sample variance will scale roughly as 
$\bS\propto1/\Nfields$ \citep[e.g.,][]{newman2002a};
however, the actual gain depends on the shape
of the \LCDM power spectrum through the survey volume
and geometry.  For a fixed amount of observing time, 
splitting a survey into $\Nfields=2$ fields will rescale 
the volume of each field by $V\propto1/\Nfields$ and
result in a corresponding increase in the typical 
dark matter density fluctuations in each field.  For
very large surveys, the decrease in the volume per
field can (at least partially) offset the gains 
achieved by probing multiple independent samples
\citep[see also][]{munoz2009a}.

\begin{figure*}
\figurenum{1}
\epsscale{1}
\plotone{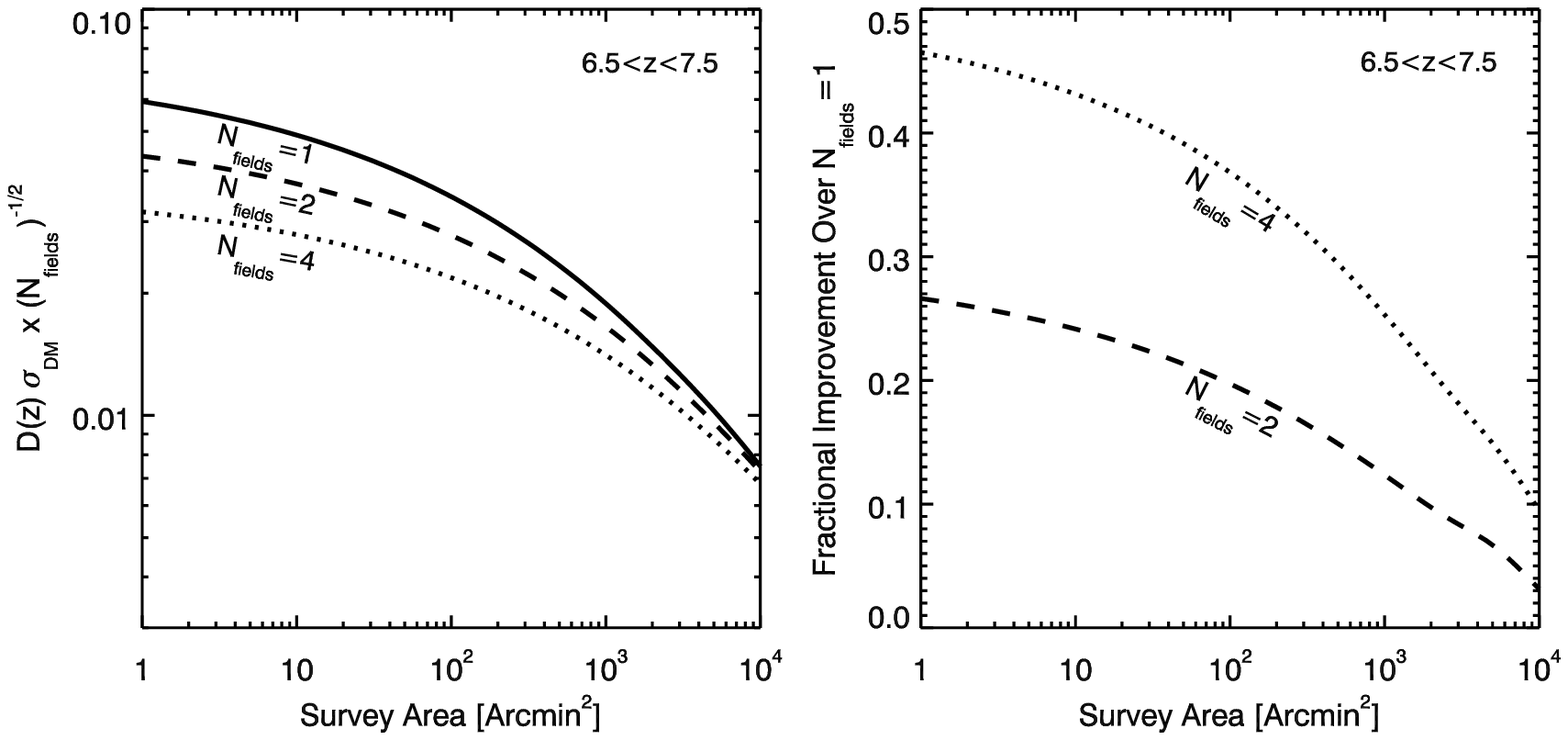}
\caption{\label{fig:multiple_fields}
Root-mean-squared (RMS) density fluctuations in a survey at redshift
$6.5 \leq z \leq 7.5$ as a function of total area (right panel).  Uncertainties 
owing to RMS density fluctuations scale with the product of the growth function $D(z)$ and
the $z=0$ RMS dark matter fluctuations $\sigmaDM$.  
If the \LCDM power spectrum $\sigmaDM$ for the rectangular survey geometry was flat, multiple field surveys
would improve their combined RMS density fluctuations by $1/\sqrt{\Nfields}$
if they probed widely-separated, statistically-independent regions.
However, the shape of the \LCDM power spectrum $\sigmaDM$ limits this
improvement since the multiple fields each probe a volume $V/\Nfields$ and
the power increases toward small scales.
Shown are the RMS density fluctuations for multiple field surveys 
($\Nfields=1$, solid line; $\Nfields=2$, dashed line; $\Nfields=4$, dotted line).
The left panel shows the fractional improvement in uncertainties owing to
density fluctuations gained by splitting survey into $\Nfields=2$ (dashed line)
or $\Nfields=4$ smaller fields of equivalent total area.  If $\sigmaDM$ were
independent of scale, the fractional improvement would be a constant 
$1-1/\sqrt{2}=0.293$ for $\Nfields=2$ and $1-1/\sqrt{4}=0.5$ for $\Nfields=4$.
}
\end{figure*}

The left panel of Figure \ref{fig:multiple_fields} shows the 
RMS density fluctuations $D(z)\sigmaDM/\sqrt{\Nfields}$ in a 
survey at redshifts
$6.5\leq z\leq7.5$ as a function of total
area for multiple fields ($\Nfields=1,2,4$).
For a galaxy survey, the sample variance in each luminosity 
bin will be increased by a factor of the galaxy bias (see Equation \ref{eqn:sample_covariance}).
While the uncertainty from RMS density fluctuations will improve with the addition of
statistically-independent samples, the fractional
improvement is less than $1-1/\sqrt{\Nfields}$
for large volumes.  The right panel of Figure \ref{fig:multiple_fields}
shows the fractional improvement gained by multiple 
fields.  For a flat power spectrum, the improvement would
be $1-1/\sqrt{2}=0.293$ for $\Nfields=2$ and $1-1/\sqrt{4}=0.5$
for $\Nfields=4$.

%%%%%%%%%%%%%%%%%%%%%%%%%%%%%%%%%%%%%%
%
%	Poisson Variance from Galaxy Abundance 
%
%%%%%%%%%%%%%%%%%%%%%%%%%%%%%%%%%%%%%%
\subsection{Poisson Variance from Galaxy Abundance}
\label{subsection:poisson_covariance}

Number-counting statistics will naturally introduce
a Poisson variance into the galaxy number count statistics.
The diagonal Poisson variance matrix $\bP$ will only add to the total covariance
for counts within a single magnitude bin (i.e., only when
$i=j$).  For definiteness, we will express the Poisson
covariance as
\begin{equation}
\label{eqn:poisson_variance}
P_{ij} = \frac{\delta_{ij} \nbar_{i}}{V_{i}}
\end{equation}
\noindent
where the Kronecker $\delta_{ij}=1$ for $i=j$ and
$\delta_{ij}=0$ for $i\ne j$.

%%%%%%%%%%%%%%%%%%%%%%%%%%%%%%%%%%%%%%
%
%	Parameter Estimation and the Fisher Matrix 
%
%%%%%%%%%%%%%%%%%%%%%%%%%%%%%%%%%%%%%%
\section{Parameter Estimation and the Fisher Matrix}
\label{section:fisher_matrix}

\citet{fisher1935a} illustrated how to infer inductively 
the properties of statistical populations from
data samples.  By approximating the likelihood function as
a Gaussian near its maximum and assuming a parameterized
model, the uncertainties in the model parameters allowed
by a future data set can be estimated directly from the data 
covariances.  Interested readers should refer to the 
excellent and detailed discussion of the Fisher matrix
formalism provided in \S 2 of \cite{tegmark1997a}, but
we outline the general approach below.

We aim to estimate the uncertainties on model parameters $\pa$
(the ``parameter covariance matrix'' $\bC$) achieved by
the data produced by some fiducial survey.  The quality of
the future data for each fiducial survey will be characterized by
the ``data covariance matrix'' $\bD$. 
The elements $\Dij$ of the total data covariance matrix 
are simply the sum of the sample covariance and Poisson
uncertainties described in \S \ref{subsection:sample_covariance}
and \S \ref{subsection:poisson_covariance}, which we
can write as
\begin{equation}
\label{eqn:data_covariance_matrix}
\Dij = \Sij + \Pij,
\end{equation}
\noindent
where the $\Pij$ only contribute when $i=j$ (see Equation \ref{eqn:poisson_variance}).
Following \cite{lima2004a}, who applied the Fisher matrix approach
to the parameter estimation of the mass--observable relation in galaxy cluster surveys,
we will express our approximate Fisher matrix as
\begin{equation}
\label{eqn:fisher_matrix}
\Fab = \sum_{ij} \frac{\pd \nbar_{i} }{\pd \pa} \left(\bD^{-1}\right)_{ij} \frac{\pd \nbar_{j} }{\pd \pb} + \frac{1}{2} \Tr\left[\bD^{-1}\frac{\pd \bS}{\pd \pa}\bD^{-1}\frac{\pd \bS}{\pd \pb}\right]
\end{equation}
\noindent
\citep[see, e.g.,][and especially the discussion in 
\S III of \citealt{lima2004a}]{holder2001a, hu2003a, lima2005a, hu2006a, cunha2009a, wu2009a}.
Here, $\Tr(\mathbf{A}) = \sum_{i=1}^{m} A_{ii}$ for an $m\times m$ matrix.
The vector elements $\pa$ reflect the parameters of the data model $\nbar$.
The first term of Equation \ref{eqn:fisher_matrix} models second derivatives of the
likelihood function in the Poisson error-dominated regime 
\citep{holder2001a}, while the second term models the sample covariance-dominated
regime \citep[][see also Appendix A of \citealt{vogeley1996a}]{tegmark1997a}.
The derivatives $\pd \nbar / \pd \pa$ of the luminosity function model 
are computed directly by differentiating Equation \ref{eqn:luminosity_function}.
The derivatives $\pd \bS/ \pd \pa$ of the sample covariance matrix are 
evaluated numerically since changes to the model luminosity function alter
the galaxy bias $b$ in Equation \ref{eqn:sample_covariance} for a given luminosity
bin in a nontrivial way.

Once the Fisher matrix $\bF$ is calculated, estimating the
parameter covariance matrix $\bC$ becomes straightforward.
The elements of the parameter covariance matrix are
approximated as
\begin{equation}
\label{eqn:parameter_covariance_matrix}
\Cab \approx (\bF^{-1})_{\mu\nu}.
\end{equation}
\noindent
The marginalized uncertainty on parameter $\pa$
is then
\begin{equation}
\label{eqn:marginalized_error}
\sigma_{\mu} \equiv C_{\mu\mu}^{1/2} = (\bF^{-1})_{\mu\mu}^{1/2}.
\end{equation}
\noindent
Similarly, we can estimate the unmarginalized error on
each parameter as $\sigma_{\mu}^{\mathrm{u}} = F_{\mu\mu}^{-1/2}$.
However, in what follows when we discuss the ``error'' or ``uncertainty'' on 
luminosity function parameters we mean the marginalized error unless
otherwise stated.

We will apply the above formalism to estimate the relative
constraining power of possible galaxy surveys, but we will
focus on evaluating such surveys in the context of existing
and forthcoming data from observational programs already
underway (i.e., the WFC3 \UDFGO~and \GOODSERS~data).  Our
statistical formalism provides a simple way to incorporate
constraints from prior data.  The combined constraints $\bCc$ of
a prior observation $\bCp$ supplemented by the forecasted constraints
of a future survey $\bC$ can be estimated as
\begin{equation}
\label{eqn:prior_knowledge}
\bCc = \left( \bC^{-1} + \bCp^{-1} \right)^{-1},
\end{equation}
\noindent
or, in other words, the combined parameter covariance
matrix is the inverse of the sum of the Fisher matrices
of the prior and future surveys.  Depending on the
magnitude of the off diagonal elements of $\bC$ and
$\bCp$, the combined parameter covariance matrix $\bCc$ can
provide a substantially different correlation between
parameters than either the prior or future surveys
produce individually.  As a result, the marginalized
uncertainty on parameters can benefit substantially
by combining surveys with different characteristics.
These ramifications of Equation \ref{eqn:prior_knowledge} will
become more apparent in \S \ref{section:results}.

%%%%%%%%%%%%%%%%%%%%%%%%%%%%%%%%%%%%%%
%
%	Model for Survey Data
%
%%%%%%%%%%%%%%%%%%%%%%%%%%%%%%%%%%%%%%
\section{Model for Survey Data}
\label{section:surveys}

Our statistical formalism for forecasting
constraints on properties of the galaxy 
population requires a model for the survey
data.  Given the approach outlined in \S 
\ref{section:variances}, the relevant
characteristics of each survey include
the total area $\Atot$, number of fields
$\Nfields$, and the minimum and maximum 
redshifts of the survey $\zmin$ and 
$\zmax$ (that, in combination with $\Atot$,
determine the survey volume $V$).  The
limiting magnitude depth of the survey
$\Mmax$ strongly influences
source statistics of the survey by determining
the faintest luminosity bin calculated
via Equation \ref{eqn:luminosity_function}.
The completeness of the survey $\fcomp$
and
halo occupation fraction $\focc$
change the cosmic sample variance by
altering the halo-galaxy correspondence in
Equation \ref{eqn:abundance_matching}. 

Of these survey characteristics, we will
keep $\zmin$, $\zmax$, $\fcomp$, and 
$\focc$ fixed between surveys.  We will
assume that the surveys are effectively
volume-limited ($\fcomp=1$) over the
redshift range of interest.  Given
a complete volume-limited survey, the
choice of minimum and maximum redshifts
roughly corresponds to the filter choice
defining a dropout selection.  We will
adopt $\zmin=6.5$ and $\zmax=7.5$, which
roughly approximates the redshift selection
of the ($z_{850}$-$Y_{105}$) vs. 
($Y_{105}$-$J_{125}$) color selection of
\citet[][see their Fig. 1]{oesch2010a}.
Similar selections can be defined for 
$I_{814}$-dropouts.  Our calculations
can be easily extended to different
redshift selections, but we adopt this
redshift range since the fiducial abundance 
of $z\sim7$ WFC3 \UDFGO~galaxy candidates 
appears increasingly robust 
\citep[see, e.g., the discussion in \S 2 of][]{mclure2009a},
the characteristic ultraviolet (UV) luminosity
of galaxies is decreasing with redshift 
\citep[e.g.,][]{bouwens2008a}, and the abundance
of dark matter halos hosting galaxies is rapidly
declining at earlier epochs.

Our
model surveys will consist of WFC3 $H$-band 
coverage with equal coverage in an additional, 
bluer WFC3 filter.  The existing
and ongoing \UDFGO~and \GOODSERS~surveys will
use the $\FSIXTY$ band (see 
\S\ref{section:existing_surveys} below),
but using the $\FFOURTY$ band buys $\approx0.3-0.5$
magnitudes in sensitivity for the same exposure
time, depending on the source luminosity (see below).
While the dropout color selection is perhaps better for
$\FSIXTY$, we will assume in our forecasts that
future surveys will utilize $\FFOURTY$. 
Our results would be similar for $\FSIXTY$
surveys to similar limiting depths.
We will characterize the abundance of
high-redshift galaxies in terms of a
the rest-frame ultraviolet (UV) luminosity
function.  We must therefore adopt a color
conversion between $\HAB$ magnitude and 
rest-frame luminosity appropriate for $z\sim7$.
In an approximation to the conversion used by
\cite{oesch2010a}, we estimate that $\HAB\approx29.0$
translates to $\MUV\approx-18.2$.  We have checked that
our general conclusions about the relative constraining
power of survey designs are insensitive to changes in this 
conversion (e.g., $\pm0.4$ magnitudes in $\MUV$).

Lastly, HST observations are 
conducted using some number $\Norbits$
per pointing that effectively determines $\Mmax$.
During each orbit the field visibility depends
on the field declination,
and the available on-source integration time
also depends on 
observatory
and instrument overheads such as guide star
acquisition, filter changes, dithering, and 
readout.  The \UDFGO~and
\GOODSERS~surveys are at a declination of 
$\delta\approx-27\deg$,
and for ease of comparison we will assume 
all future surveys have $|\delta|<30\deg$.
This roughly equatorial declination range provides
a visibility of 54 minutes/orbit\footnote{{\scriptsize see Table 6.1 of
http://www.stsci.edu/hst/proposing/documents/primer/}}.
Given the additional observatory and instrument overheads,
we will calculate all sensitivities using a 46 minutes/orbit 
exposure time.  Given the compact character of the observed
WFC3 $z\sim7$ galaxy candidates \citep[e.g.,][]{oesch2010b}, we will report
optimum 5-$\sigma$ point source sensitivities.
Table \ref{table:sensitivity} lists these
sensitivities for a flat
$F_{\nu}$ spectrum source, 
as a function of $\Norbits$ for
both the $\FFOURTY$ and $\FSIXTY$ filters\footnote{\scriptsize Computed using
the WFC3 IR Channel Exposure Time Calculator, 
http://etc.stsci.edu/webetc/mainPages/wfc3IRImagingETC.jsp.}.

\begin{deluxetable}{ccc}
\tablecolumns{3} 
\tablewidth{0pc} 
\tablecaption{Optimum 5-$\sigma$ Point Source Sensitivity vs. Exposure Time Per Pointing}
\tablehead{
\colhead{$\Norbits$} & \colhead{$H_{140}$ [AB Mag.]} & \colhead{$H_{160}$ [AB Mag.]}
\label{table:sensitivity}}
\startdata 
0.5  & 27.00 & 26.62 \\
1.0  & 27.43 & 27.07 \\
2.0  & 27.84 & 27.49 \\
3.0  & 28.07 & 27.73 \\
4.0  & 28.23 & 27.89 \\
6.0  & 28.46 & 28.12 \\
8.0  & 28.62 & 28.28 \\
19.0 & 29.10 & 28.76 \\
38.0 & 29.47 & 29.14 \\
125.0 & 30.12 & 29.78
\enddata 
\end{deluxetable}

%%%%%%%%%%%%%%%%%%%%%%%%%%%%%%%%%%%%%%
%
%	Existing Surveys
%
%%%%%%%%%%%%%%%%%%%%%%%%%%%%%%%%%%%%%%
\section{Existing Surveys}
\label{section:existing_surveys}

The discussion in \S \ref{section:variances} makes clear
that the combination of different survey designs can 
potentially provide increased constraints beyond that
achieved by individual data sets.  Even duplicate
surveys will reduce the Poisson variance and potentially
the sample variance (especially if the fields are widely
separated on the sky).  In the absence of significant
systematic biases, using prior data will generally improve
the expected parameter uncertainty obtained by future
experiments.  We will therefore rely on the expected
constraints achieved by the ongoing WFC3 \UDFGO~(PI 
Illingworth, Program ID 11563) and 
\GOODSERS~(PI O'Connell, Program ID 11359) programs 
to augment the fiducial survey designs
evaluated in \S \ref{section:model_surveys} and 
\S \ref{section:results}.  In this section, we
will calculate the expected constraints provided
by the \UDFGO~and \GOODSERS~data.

\begin{deluxetable*}{lcccccc}
\tablecolumns{6} 
\tablewidth{0pc} 
\tablecaption{Existing Surveys} 
\tablehead{
\colhead{Survey} & \colhead{Field Geometry}            & \colhead{$\Nfields$} & \colhead{Total Area}    & \colhead{$\Norbits$\tablenotemark{a}} & \colhead{$H$-band Depth} & \colhead{Ref.}\\
\colhead{Name}  & \colhead{[Arcmin. $\times$ Arcmin.]} & \colhead{[\#]}       & \colhead{[Sq. Arcmin.]} & \colhead{[\#]} & \colhead{[AB Mag.]} &
\label{table:existing_surveys}}
\startdata 
\UDFGO   & 2.05\arcmin  $\times$ 2.27\arcmin & 2 & 9.3  &19& 28.76 & 1\\
         & 2.05\arcmin  $\times$ 2.27\arcmin & 1 & 4.7  &38& 29.14 & \\
\GOODSERS & 5.2\arcmin  $\times$ 10.3\arcmin\tablenotemark{b} & 1 & 53.3 &3 & 27.73\tablenotemark{c} & 2
\enddata 
\tablenotetext{a}{Number of orbits per pointing.}
\tablenotetext{b}{The \GOODSERS~survey is a $2\times4$ WFC3 IR mosaic dithered to match the Ultraviolet/Visible channel field of view.}
\tablenotetext{c}{For details, see the discussion in \S \ref{section:existing_surveys}.}
\tablerefs{ (1) http://www.stsci.edu/observing/phase2-public/11563.pdf; (2) http://www.stsci.edu/hst/proposing/old-proposing-files/goods-cdfs.pdf
}
\end{deluxetable*}

Table \ref{table:existing_surveys} describes the
field geometry, number of fields $\Nfields$, total
area, and expected $H$-band $5-\sigma$ point source
depth for the \UDFGO~and \GOODSERS~survey designs.
Numerous analyses of the initial \UDFGO~data release
have already been performed 
\citep[e.g.,][]{bouwens2010b,oesch2010a,oesch2010b,
mclure2009a,bunker2009a,mclure2009a,yan2009a,finkelstein2009a},
but we will consider the expected constraints 
provided by the entire 192 orbit program.  
The
\GOODSERS~data has not yet been released 
\citep[but for initial analyses on unreleased ERS data see][]{wilkins2009a,labbe2009a}, and
we will also use the Fisher matrix approach to
estimate the constraints provided by that survey. 

The \UDFGO~program is comprised of three WFC3 pointings.
Two of the WFC3 pointings use the \FSIXTY~filter with 
19 orbits.
Using the WFC3 ETC, we estimate these observations will
reach $\HAB\approx28.76$.
The \UDFGO~observations also will have 
38 \FSIXTY~orbits in the HUDF that will reach 
$\HAB\approx29.14$\footnote{\scriptsize The \UDFGO~proposal estimates
their sensitivities as $\HAB\approx28.6$ for 19 orbits and
$\HAB\approx29$ for 38 orbits.  See http://www.stsci.edu/observing/phase2-public/11563.pdf.}.
The remaining 116 orbits in the program will be used 
for observing in bluer filters.  

The
\GOODSERS~survey will have 3-orbit depth in \FSIXTY,
using 24 orbits (out of a total 104) for $H$-band observations.
We estimate that these observations will reach
a sensitivity of $\HAB\approx27.73$\footnote{\scriptsize The \GOODSERS~proposal
estimates the 3-orbit sensitivity using 40 minutes/orbit integration
as $\HAB\approx26.5$.  
See http://www.stsci.edu/hst/proposing/old-proposing-files/goods-cdfs.pdf
and http://www.stsci.edu/observing/phase2-public/11359.pdf.
Our method for estimating the sensitivity
would provide $\HAB\approx27.65$ for 40 minutes/orbit.}  
The remaining 80 orbits in the program will be used for
observations with other filters and grisms.

\begin{figure*}
\figurenum{2}
\epsscale{1.1}
\plotone{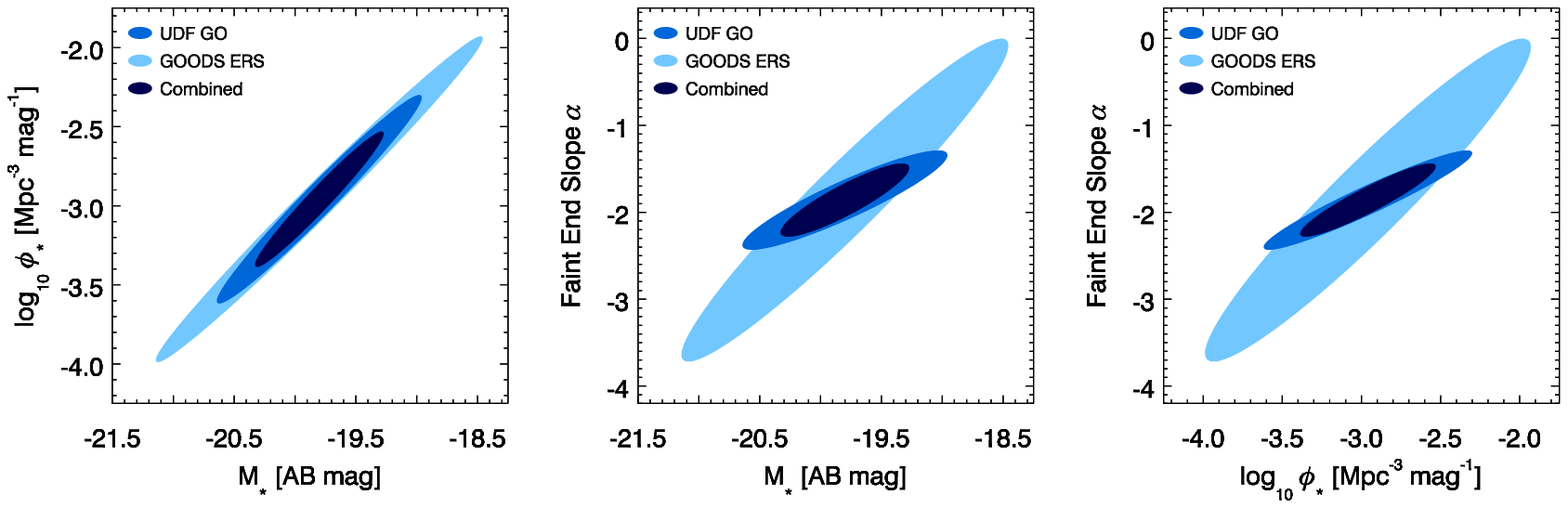}
\caption{\label{fig:survey_udf_go_goods_ers}
Forecasted constraints on $z\sim7$ luminosity function parameters expected from the forthcoming \UDFGO~and \GOODSERS~WFC3 data.
The constraints are calculated for a \citet{schechter1976a} luminosity function with a characteristic luminosity $\Mstar$, normalization
$\phistar$, and faint-end slope $\alpha$.
Shown are the $1-\sigma$ constraints in the $\Mstar-\logten \phistar$ (left panel), $\Mstar-\alpha$ (middle panel), and $\logten \phistar-\alpha$ (right 
panel) space projections for the \UDFGO~(blue region) and \GOODSERS~(light blue region) surveys.  Also shown are the constraints
expected by combining both surveys (dark blue region).  The depth of the \UDFGO~survey will provide a better constraint on the
faint-end slope than \GOODSERS, but the differences between their parameter covariances make them complementary.
}
\end{figure*} 

\subsection{Forecasted Constraints for Existing Surveys}
\label{subsection:constraints_from_existing_surveys}

The forecasted constraints achieved by the \UDFGO~and \GOODSERS~surveys
are plotted in Figure \ref{fig:survey_udf_go_goods_ers}.  Each panel
shows the constraints for the \UDFGO~(blue region) and \GOODSERS~(light blue region)
surveys separately, and in combination (dark blue region). 
The constraints are plotted for the $\Mstar$-$\phistar$ (left panel),  $\Mstar$-$\alpha$ (middle panel), 
and $\phistar$-$\alpha$ (right panel) projections.
In Figure \ref{fig:survey_udf_go_goods_ers} (and 
in similar figures throughout the paper), the shaded regions correspond 
to standard Gaussian contours.

Figure \ref{fig:survey_udf_go_goods_ers} highlights some general
properties of the performance of different kinds of surveys for
providing luminosity function parameter constraints, as
well as specific features of the \UDFGO~and \GOODSERS~surveys:
\begin{itemize}
\item The covariances between luminosity function parameters are significant and positive.
In terms of the Pearson's correlation coefficient $\rho$ for parameters $x$ and $y$,
defined as
\begin{equation}
\label{eqn:pearsons_correlation_coefficient}
\rho = \frac{C_{xy}}{\sigma_{x}\sigma_{y}},
\end{equation}
\noindent
the forecasts calculate typical correlation coefficients of $\rho\gtrsim0.9$.
For a given $\Mstar$ a narrow range of $\logten\phistar$ or
$\alpha$ are permitted by the data, even as the marginalized uncertainties
can be as large as $\sim30-40\%$ fractionally for $\logten\phistar$ and $\alpha$.  

\item The orientation
of the constraint ellipse forecasted for each survey can 
differ significantly depending on the parameter uncertainties,
even if the parameter correlation coefficients for the separate surveys are similar.
The orientation of the constraint ellipse major axis with respect to the
$x$-axis in the $x$-$y$ parameter plane can be characterized by
the angle
\begin{equation}
\label{eqn:orientation_angle}
\oangle \equiv \frac{1}{2}\arctan\left[\frac{2\rho\sigma_{x}\sigma_{y}}{\sigma_{x}^{2}-\sigma_{y}^{2}}\right] = \frac{1}{2}\arctan\left[\frac{2C_{xy}}{C_{xx}-C_{yy}}\right],
\end{equation}
\noindent
which depends on the correlation $\rho$ and the parameter uncertainties.
If the angle $\oangle$ differs between separate surveys, then the
constraints achieved by combining the surveys can improve dramatically.
\end{itemize}
\noindent
For reference, the calculated marginalized and unmarginalized errors 
for $\bp$ as well as the 
Pearson's correlation coefficient $\rho$ and the angle $\oangle$ 
for each pair of
parameters are listed for the existing surveys 
in Table \ref{table:existing_survey_constraints}.

As Figure \ref{fig:survey_udf_go_goods_ers} and Table \ref{table:existing_survey_constraints}
show, the \UDFGO~and \GOODSERS~surveys will already provide interesting
constraints on the abundance of $z\sim7$ galaxies.  When combined, the unmarginalized 
uncertainties on the LF parameters will be $\Delta\Mstar\sim0.1$ mag,
$\Delta\logten\phistar\lesssim0.1$, and $\Delta\alpha\sim0.15$.
For the full \UDFGO~survey, we find that the ummarginalized uncertainty for the 
faint-end slope is $\Delta\alpha\sim0.16$.  Using a single \UDFGO~field and a limiting depth of
$\FSIXTY\sim29$ AB for a single pointing, \cite{oesch2010a} report a faint-end slope
uncertainty of $\Delta\alpha\sim0.33$ when $\phistar$ and $\Mstar$ are fixed (i.e., the
unmarginalized uncertainty on $\alpha$).  If we use the same single pointing area
and depth, and the same cosmology, our estimate of the unmarginalized uncertainty would 
increase to $\Delta\alpha\sim0.26$.
The \GOODSERS~and \UDFGO~surveys
are complementary in that the depth of the \UDFGO~survey provides a beneficial
constraint on the faint-end slope $\alpha$. This \UDFGO~constraint
on $\alpha$ rotates the \UDFGO~error ellipse relative to the \GOODSERS~constraint
in the $\Mstar-\alpha$ and $\phistar-\alpha$ projections, thereby reducing
the corresponding parameter uncertainties.  Individually, the 
\GOODSERS~uncertainties will be considerably larger than those obtained by the 
\UDFGO~survey, since the \GOODSERS~survey lacks sufficient depth to tightly
constrain the LF faint-end slope and is not wide enough to tightly
constrain $\Mstar$ or $\phistar$.

While the \UDFGO~and 
\GOODSERS~surveys achieve appreciable
unmarginalized constraints,
the covariances between
the LF parameters are large. The marginalized uncertainties calculated
for the
LF parameters are $\Delta\Mstar\sim0.5$ mag, $\Delta\phistar\sim0.4$,
and $\Delta\alpha\sim0.4$.
Accounting for covariances these marginalized parameter uncertainties 
correspond to a fractional uncertainty in the total number of galaxies 
with $\MUV<-18$ of $\approx25\%$, increasing to a factor of $\approx2$ uncertainty
in the total number of galaxies with $\MUV<\Mstar$.
To improve the constraints on the number of galaxies with 
$\MUV<-18$ ($\MUV<\Mstar)$ to $\approx5\%$ ($\approx50\%$) would require
marginalized parameter uncertainties of approximately
$\Delta\Mstar\sim0.2$ mag, $\Delta\phistar\sim0.2$,
and $\Delta\alpha\sim0.2$ depending on their covariances.
To reach such constraints, these \UDFGO~and \GOODSERS~surveys 
would need to be complemented by either wider area or deeper surveys. 
We now consider some fiducial model surveys that could achieve
these constraints in combination with the \UDFGO~and \GOODSERS~data.

%%%%%%%%%%%%%%%%%%%%%%%%%%%%%%%%%%%%%%
%
%	Model Surveys
%
%%%%%%%%%%%%%%%%%%%%%%%%%%%%%%%%%%%%%%
\section{Model Surveys}
\label{section:model_surveys}

The complete \UDFGO~and \GOODSERS~surveys
will provide extremely interesting initial data on 
the abundance of $z\sim7$ galaxies, but the
marginalized uncertainties on the LF
parameters achieved by those surveys will
still permit uncertainties of $\sim25\%$
in the total number of galaxies at 
$\MUV\lesssim-18$.  We can repeat the
calculations from \S \ref{section:existing_surveys}
for 
fiducial model surveys to illustrate
what constraints wider or deeper surveys 
can achieve when combined with the
\UDFGO~and \GOODSERS~data.

The model surveys are designed to be
appropriate for a HST Multi-Cycle Treasury
Program\footnote{See, e.g., http://www.stsci.edu/institute/org/spd/mctp.html/},
which can receive up to 750
orbits per HST cycle\footnote{See http://www.stsci.edu/institute/org/spd/HST-multi-cycle-treasury}.
We consider six possible model surveys that we 
estimate would require $450-900$ total orbits
to acquire filter coverage with the WFC IR channel.

As discussed in \S \ref{section:surveys}, we will assume
the model surveys will use the $\FFOURTY$ filter owing to
its increased throughput relative to $\FSIXTY$.  The
sensitivity of each survey is determined by selecting 
a number $\Norbits$ of orbits per pointing, assuming 46 
minutes/orbit exposure time, and using the WFC3 IR
channel ETC.  The total number of orbits for each
survey were then determined by selecting the number of
fields $\Nfields$, a mosaic
geometry per field, and multiplying the number of pointings in each
mosaic by $\Norbits$ (and then doubling to account for
comparable coverage in a bluer WFC3 filter).

The survey models are designed to cover a large range in
total area ($\Atot\approx14-3600$ square arcmin),
field numbers ($\Nfields=1-4$), 
orbits per pointing ($\Norbits=0.5-125$), limiting 
depth ($\HAB\approx27-30$),
and 
total number of orbits ($450-900$).  We design each survey
to approximate possible HST WFC3 tilings of existing
surveys; as such, these model surveys represent realistic
extensions of existing HST and Spitzer surveys to hundreds
of orbits of WFC3 coverage.  Summaries of the model surveys
can be found in Table \ref{table:model_surveys}, and are
ordered by decreasing limiting depth and increasing total area.
Brief descriptions of the models follow:

\begin{deluxetable*}{lccccccc}
\tablecolumns{8} 
\tablewidth{0pc} 
\tablecaption{Model Surveys} 
\tablehead{
\colhead{Model} & \colhead{Field Mosaic} & \colhead{Field Geometry} & \colhead{$\Nfields$} & \colhead{Total Area} & \colhead{$\Norbits$\tablenotemark{a}} & \colhead{$H$-band Depth} & \colhead{Total Orbits\tablenotemark{b}}\\
\colhead{Name}  & \colhead{[\# Point. $\times$ \# Point.]} & \colhead{[Arcmin. $\times$ Arcmin.]} & \colhead{[\#]}       & \colhead{[Sq. Arcmin.]} & \colhead{[\#]} & \colhead{[AB Mag.]} & \colhead{[\#]}
\label{table:model_surveys}}
\startdata 
\UDFDeep      & $1\times1$   &  2.05\arcmin $\times$ 2.27\arcmin & 3 & 13.96  & 125   & 30.12 & 750\\
\GOODSDeep    & $5\times7$   & 10.3\arcmin  $\times$ 15.9\arcmin & 2 & 325.7  &   8   & 28.62 & 896\\
\GOODSMedium  & $5\times7$   & 10.3\arcmin  $\times$ 15.9\arcmin & 2 & 325.7  &   6   & 28.46 & 672\\
\GOODSShallow & $5\times7$   & 10.3\arcmin  $\times$ 15.9\arcmin & 2 & 325.7  &   4   & 28.23 & 448\\
\SEDS         & \ $4\times13$  & \ \ 8.2\arcmin  $\times$ 29.5\arcmin & 4 & 967.6  &   2   & 27.84 & 832\\
\COSMOS       & $26\times30$ & 59.0\arcmin  $\times$ 61.5\arcmin & 1 & 3628.5 &   0.5 & 27.00 & 780
\enddata 
\tablenotetext{a}{Number of orbits per pointing.}
\tablenotetext{b}{We assume each survey will require comparable coverage in two WFC3 filters, which
doubles the required number of total orbits.
}
\end{deluxetable*}

\subsection{\UDFDeep}  The performance of the \UDFGO~survey suggests
that an interesting possible survey design would be a set of
narrow pencil beam surveys with sufficient depth to reach
a few nJy sensitivity.  Extending each of the three single WFC3 pointing \UDFGO~fields
to $\sim125$ orbits in \FFOURTY~would achieve $\HAB\approx30.1$\footnote{Surveys
this deep require many WFC3 frame exposures to avoid image persistence.  We ignore
the impact of any additional related overhead on the available exposure time.}.
For surveys of $\sim10$ square arcmin total area, using $\Nfields>1$ results
in a substantial reduction of sample variance (see Figure \ref{fig:multiple_fields}).
The model \UDFDeep~will therefore use $\Nfields=3$, $\Norbits=125$,
and $\Atot=13.96$ square arcmin (three WFC3 pointings), and $750$ total orbits including
coverage in a bluer WFC3 filter.  For calculating constraints from a combination of
\UDFDeep~with existing data, we assume the \UDFDeep~fields will be able to leverage
the \GOODSERS~data but will duplicate the \UDFGO~data\footnote{The \FSIXTY~data from
\UDFGO~could be incorporated the same $\HAB$-band depth, which could potentially decrease 
the total orbits for this survey by $\sim150$.  Our general conclusions are not strongly influenced by choosing this alternative.}.

\subsection{\GOODSDeep}  Another template for a model survey is deep WFC3 coverage
of the GOODS survey fields.  A $5\times7$ WFC3 mosaic could cover a field of
size $10.3\arcmin \times 15.9\arcmin$, similar to the GOODS fields \citep{giavalisco2004a}.
Covering $\Nfields=2$ fields the size of the GOODS fields would require 70 pointings, and
would cover a total area of $\Atot=326$ square arcmin.
Using $\Norbits=8$ orbits per pointing would reach $\HAB=28.6$ in \FFOURTY, and
would require a total of 896 orbits (including coverage in a bluer WFC3 filter).
\GOODSDeep~is the most expensive survey we consider.  For calculating
combined constraints utilizing existing, we will combine \GOODSDeep~with the \UDFGO~data
but ignore the duplicated \GOODSERS~\FSIXTY data\footnote{Most of the additional constraint
achieved by combining with existing data comes from the \UDFGO~data, so this choice is not
critical for our general conclusions.  However, using the \GOODSERS~data could potentially
decrease the required orbits for a GOODS-like survey by $\sim24$ orbits.}.

\subsection{\GOODSMedium}  To gain intuition about the relative value of depth and area for
constraining high-redshift galaxy populations, we will consider variations of the GOODS-like
survey.  \GOODSMedium~is identical to \GOODSDeep~in number of fields and pointings, but would
achieve a reduced depth of $\Norbits=6$ orbits per pointing ($\HAB=28.46$).  The total number
of orbits required for \GOODSMedium~is 672 (including equal coverage in a bluer WFC3 filter).
When determining combined constraints with existing data, we will combine \GOODSMedium~with 
the \UDFGO~survey.

\subsection{\GOODSShallow}  Same as \GOODSDeep~and \GOODSMedium, but to $\Norbits=4$ orbits per pointing 
($\HAB=28.2$) depth.  \GOODSShallow~would require 448 total orbits.  For combined constraints
with existing data, we will combine \GOODSShallow~with \UDFGO.

\subsection{\SEDS}  An existing survey with a combination of large area and infrared depth is the Spitzer Extended
Deep Survey \citep[SEDS;][]{fazio2008a}, which was designed to cover $0.9\deg^{2}$ over five fields
to 12 hour/pointing depth with the warm Spitzer Infrared Array Camera $3.6\micron$ and $4.5\micron$ 
channels.  Exactly reproducing the SEDS survey with WFC3 would likely be prohibitively expensive, so
we will instead consider a feasible WFC3 survey with a design similar in spirit to SEDS.  Our SEDS-like
\SEDS~will consist of $\Nfields=4$ fields of $4\times13$ pointing mosaics (each of size 
$8.2\arcmin\times29.5\arcmin$), for a total area $\Atot=967.6$ square arcmin.  A depth of $\Norbits=2$
orbits per pointing ($\HAB=27.8$) would then require 832 orbits (including equal coverage in a bluer
WFC3 filter).  For calculating combined constraints with existing data, we will combine \SEDS~with
both the \UDFGO~and \GOODSERS~fields.

\subsection{\COSMOS}  The largest HST survey to date is the equatorial Cosmic Origins Survey \citep[COSMOS][]{scoville2007a},
which covers $2\deg^{2}$ with the ACS $I$-band.  As with the SEDS-like \SEDS, exactly reproducing the COSMOS 
survey with WFC3 would likely be prohibitively expensive.  Instead, we consider a $1\deg^{2}$ 
($\Atot=3629$ square arcmin) survey with a single $26\times30$ mosaic ($59.0\arcmin\times61.5\arcmin$) to
$\Norbits=0.5$ orbits per pointing ($\HAB=27$) depth.  \COSMOS~is the widest and shallowest design
we consider, and would require 780 orbits to complete (including equal coverage in a bluer
WFC3 filter).  We will combine \COSMOS~with both
the \UDFGO~and \GOODSERS~surveys for purposes of calculating combined constraints incorporating
existing data.

\subsection{Field Size Comparison}
\begin{figure*}
\figurenum{3}
\epsscale{0.8}
\plotone{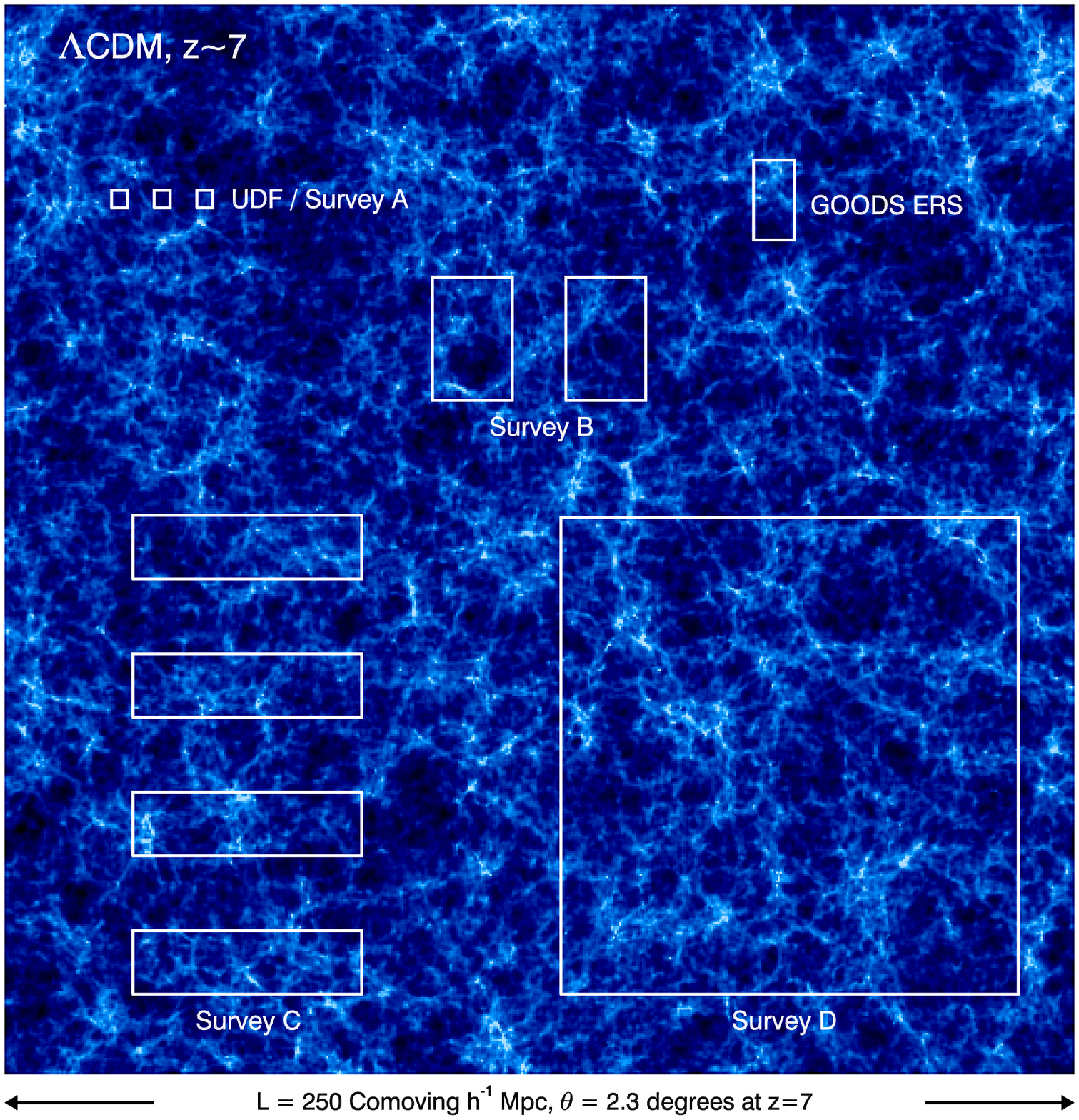}
\caption{\label{fig:survey_areas}
Existing and model survey areas compared with the large scale dark matter structure at $z\sim7$.
Shown are the \UDFGO, \GOODSERS, \AllSurveys~areas (white boxes), projected onto a surface density
map of a thin $10h^{-1}\Mpc$ slice through a \LCDM cosmological simulation of size 
$L=250h^{-1}\Mpc$ \citep{tinker2008a}.  The number of fields and survey areas of \UDFGO~and \UDFDeep~
are identical.  This comparison illustrates the characteristic angular size of
large scale structures at $z\sim7$, as well as the survey areas required to probe representative samples of
the high-redshift dark matter density distribution.  The separation between fields is not to scale, and
model surveys incorporating different fields would likely be more widely spaced to probe statistically independent
regions on the sky.
}
\end{figure*}

We show an illustrative comparison of the existing and 
model survey areas in Figure \ref{fig:survey_areas}.
The \UDFGO, \GOODSERS, and \AllSurveys~areas are shown as
white boxes overlaid on a thin $10h^{-1}\Mpc$ slice through 
a \LCDM cosmological simulation of comoving size 
$L=250h^{-1}\Mpc$ at $z\sim7$ \citep{tinker2008a}.  The blue scale image
shows the projected dark matter surface density calculated
from the dark matter particle distribution of the simulation.
The comoving length scale corresponding to an angle of $\theta=1\deg$
at $z=7$ is $108.7h^{-1}\Mpc$ for the adopted WMAP5 cosmology.
This comparison illustrates the characteristic angular size of
large scale structures at $z\sim7$, as well as the survey areas required to probe representative samples of
the high-redshift dark matter density distribution.  The separation between fields is not to scale, and
model surveys incorporating different fields would likely be more widely spaced to probe statistically independent
regions on the sky.

%%%%%%%%%%%%%%%%%%%%%%%%%%%%%%%%%%%%%%
%
%	Forecasted Constraints for Model Surveys	
%
%%%%%%%%%%%%%%%%%%%%%%%%%%%%%%%%%%%%%%
\section{Forecasted Constraints for Model Surveys}
\label{section:results}

\begin{figure*}
\figurenum{4}
\epsscale{1.1}
\plotone{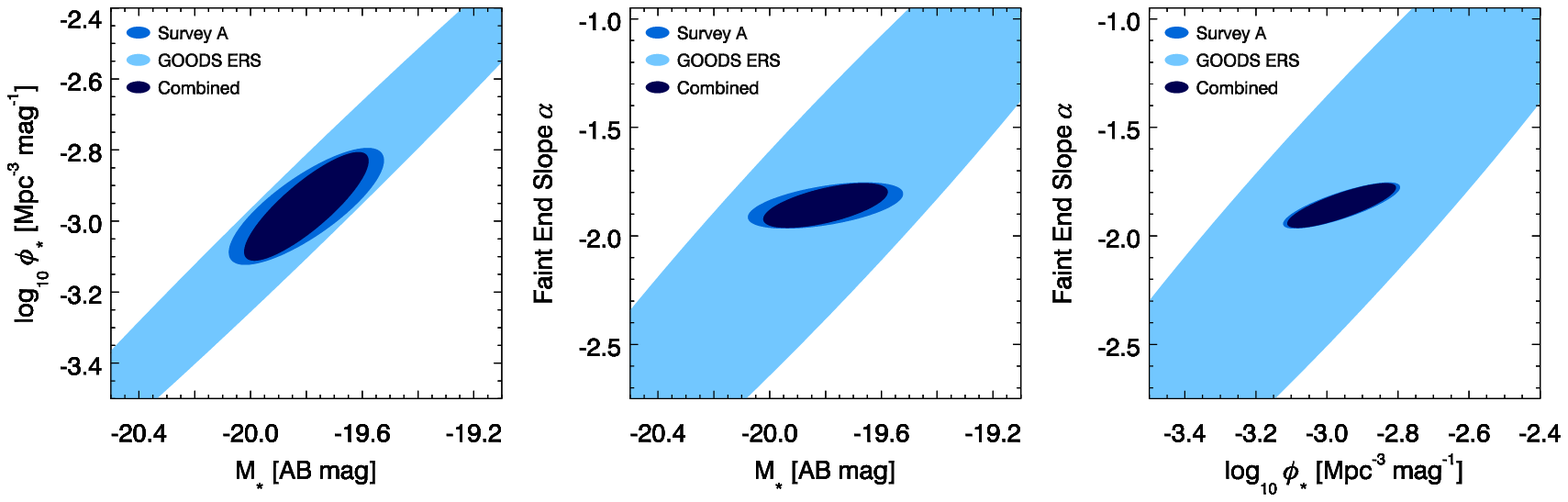}
\caption{\label{fig:survey_goods_ers_model_udf}
Forecasted constraints on $z\sim7$ luminosity function parameters expected from the existing \GOODSERS~survey and the model \UDFDeep.
The constraints are calculated for a \citet{schechter1976a} luminosity function with a characteristic luminosity $\Mstar$, normalization
$\phistar$, and faint-end slope $\alpha$.
Shown are the $1-\sigma$ constraints in the $\Mstar-\logten \phistar$ (left panel), $\Mstar-\alpha$ (middle panel), and $\logten \phistar-\alpha$ (right 
panel) space projections for the \GOODSERS~(light blue region) and \UDFDeep~(blue region) surveys.  Also shown are the constraints
expected by combining both surveys (dark blue region).  
}
\end{figure*}

\begin{figure*}
\figurenum{5}
\epsscale{1.1}
\plotone{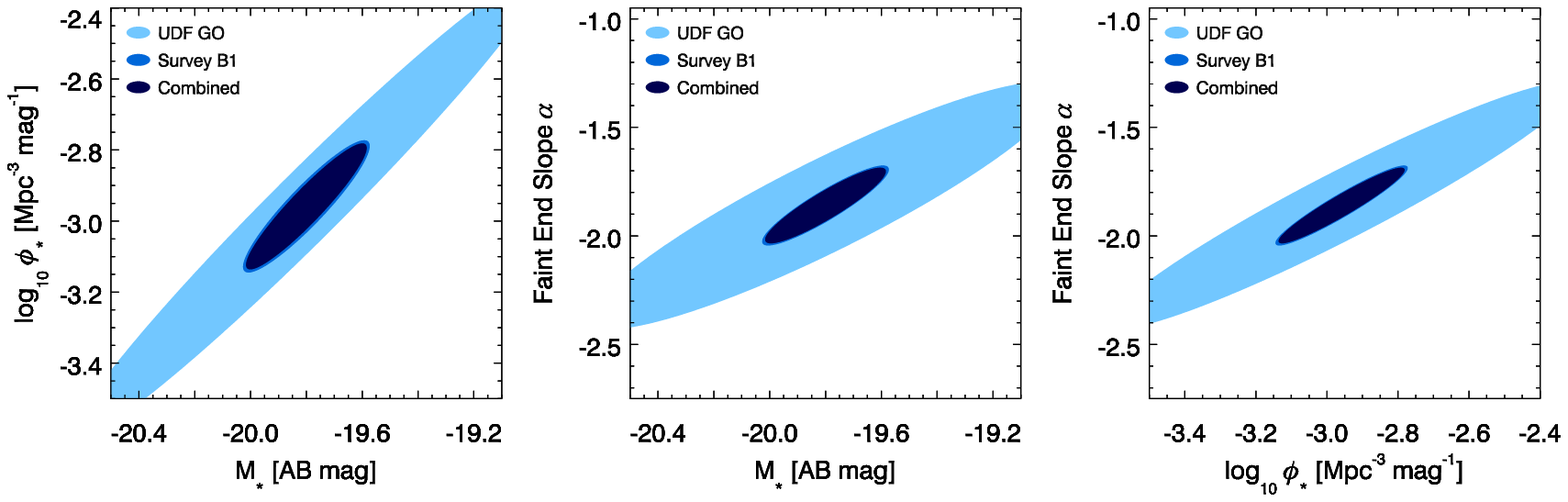}
\caption{\label{fig:survey_udf_go_model_goods_deep}
Forecasted constraints on $z\sim7$ luminosity function parameters expected from the existing \UDFGO~survey and the model \GOODSDeep.
The constraints are calculated for a \citet{schechter1976a} luminosity function with a characteristic luminosity $\Mstar$, normalization
$\phistar$, and faint-end slope $\alpha$.
Shown are the $1-\sigma$ constraints in the $\Mstar-\logten \phistar$ (left panel), $\Mstar-\alpha$ (middle panel), and $\logten \phistar-\alpha$ (right 
panel) space projections for the \UDFGO~(light blue region) and \GOODSDeep~(blue region) surveys.  Also shown are the constraints
expected by combining both surveys (dark blue region).  
}
\end{figure*}

\begin{figure*}
\figurenum{6}
\epsscale{1.1}
\plotone{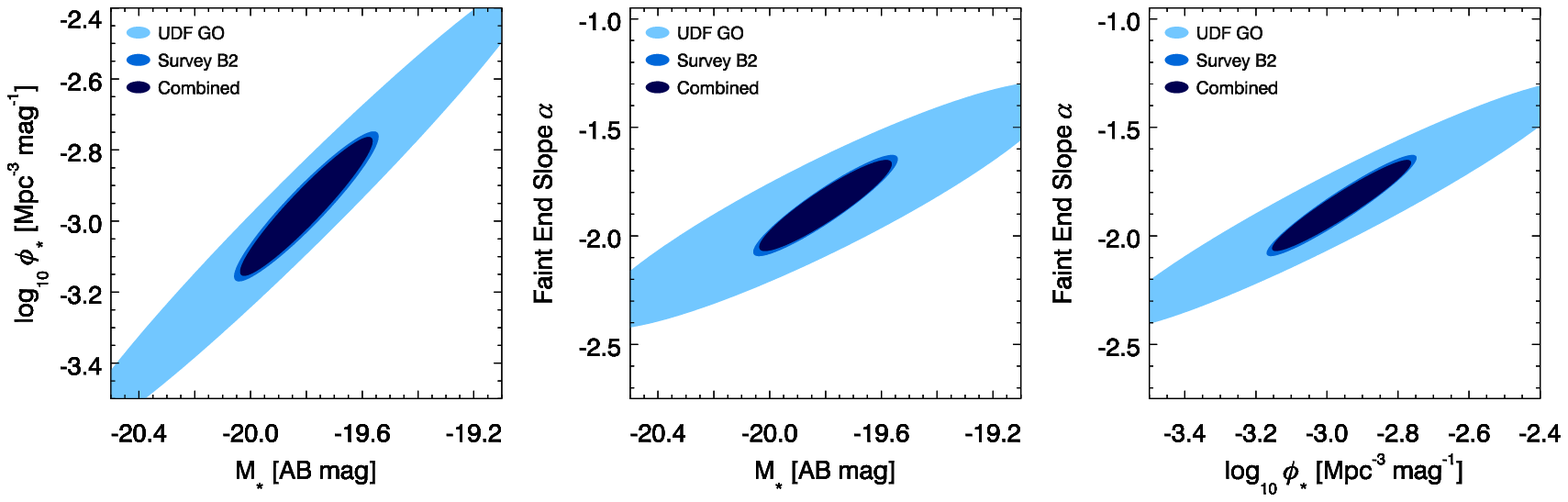}
\caption{\label{fig:survey_udf_go_model_goods_medium}
Forecasted constraints on $z\sim7$ luminosity function parameters expected from the existing \UDFGO~survey and the model \GOODSMedium.
The constraints are calculated for a \citet{schechter1976a} luminosity function with a characteristic luminosity $\Mstar$, normalization
$\phistar$, and faint-end slope $\alpha$.
Shown are the $1-\sigma$ constraints in the $\Mstar-\logten \phistar$ (left panel), $\Mstar-\alpha$ (middle panel), and $\logten \phistar-\alpha$ (right 
panel) space projections for the \UDFGO~(light blue region) and \GOODSMedium~(blue region) surveys.  Also shown are the constraints
expected by combining both surveys (dark blue region).  
}
\end{figure*}

\begin{figure*}
\figurenum{7}
\epsscale{1.1}
\plotone{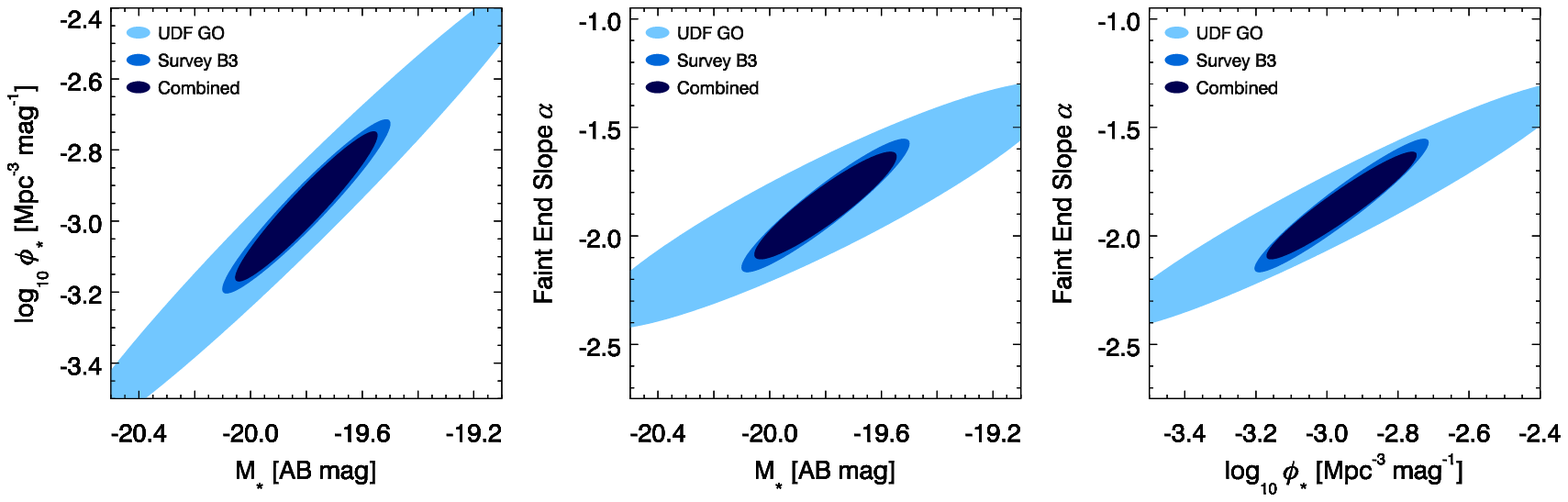}
\caption{\label{fig:survey_udf_go_model_goods_shallow}
Forecasted constraints on $z\sim7$ luminosity function parameters expected from the existing \UDFGO~survey and the model \GOODSShallow.
The constraints are calculated for a \citet{schechter1976a} luminosity function with a characteristic luminosity $\Mstar$, normalization
$\phistar$, and faint-end slope $\alpha$.
Shown are the $1-\sigma$ constraints in the $\Mstar-\logten \phistar$ (left panel), $\Mstar-\alpha$ (middle panel), and $\logten \phistar-\alpha$ (right 
panel) space projections for the \UDFGO~(light blue region) and \GOODSShallow~(blue region) surveys.  Also shown are the constraints
expected by combining both surveys (dark blue region).  
}
\end{figure*}

\begin{figure*}
\figurenum{8}
\epsscale{1.1}
\plotone{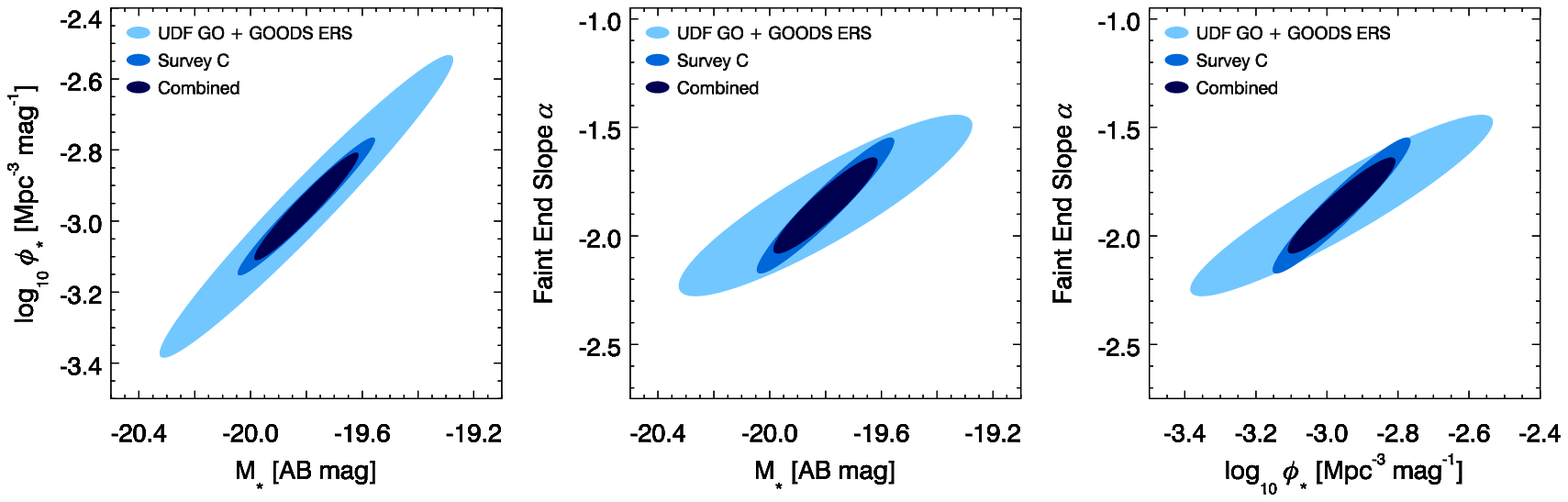}
\caption{\label{fig:survey_udf_go_goods_ers_model_seds}
Forecasted constraints on $z\sim7$ luminosity function parameters expected from the existing \UDFGO~and \GOODSERS~surveys and the model \SEDS.
The constraints are calculated for a \citet{schechter1976a} luminosity function with a characteristic luminosity $\Mstar$, normalization
$\phistar$, and faint-end slope $\alpha$.
Shown are the $1-\sigma$ constraints in the $\Mstar-\logten \phistar$ (left panel), $\Mstar-\alpha$ (middle panel), and $\logten \phistar-\alpha$ (right 
panel) space projections for the combined \UDFGO~and \GOODSERS~surveys (light blue region) and the model \SEDS~(blue region).  Also 
shown are the constraints
expected by combining all three surveys (dark blue region).  
}
\end{figure*}

The forecasted constraints calculated for the
model 
\AllSurveys~are summarized in Table \ref{table:model_survey_constraints}
and presented in 
Figures \ref{fig:survey_goods_ers_model_udf}-\ref{fig:survey_udf_go_goods_ers_model_cosmos}.
In each figure, the shaded areas show the
projected constraints for each model survey in
the $\Mstar-\phistar$ (left panel), $\Mstar-\alpha$ (middle panel), 
and $\phistar-\alpha$ (right panel)
LF parameter planes.  The axes ranges in 
Figures \ref{fig:survey_goods_ers_model_udf}-\ref{fig:survey_udf_go_goods_ers_model_cosmos}
are identical (and much smaller than in Figure \ref{fig:survey_udf_go_goods_ers}), and the
plotted constraints are directly comparable.
A description of the forecasted constraints for each model survey
follows:

\subsection{\UDFDeep}

Figure \ref{fig:survey_goods_ers_model_udf} shows 
the forecasted constraints for \UDFDeep~($\Atot=14$ square arcmin, $\Nfields=3$, $\HAB=30.1$, blue region),
\GOODSERS~(light blue region), and \GOODSERS~and \UDFDeep~combined (dark blue region).
\UDFDeep~could find $>250$ $z\sim7$ galaxies to $\HAB\sim30.1$,
with a Poisson variance in the galaxy count of $\approx6\%$..
The average galaxy bias for this depth and area is $\bave\approx5.3$,
which results in a sample cosmic variance of $\sigmaCV\approx0.16$
fractionally\footnote{For our cosmology and the \cite{oesch2010a} estimated
LF, the comoving volume of a 
single WFC3 pointing to $\HAB\sim29$ depth at the redshifts of interest would have a cosmic
variance uncertainty of $\sigmaCV\approx0.32$ \citep[c.f.,][]{oesch2010a}.}.
The constraints achieved by \UDFDeep~are therefore cosmic variance limited.
\UDFDeep~is the deepest model survey we consider, and results in the
tightest forecasted constraints on the LF faint-end slope 
($\Delta\alpha\approx0.1$, marginalized).  \GOODSERS~complements 
\UDFDeep~by providing an improved combined constraint on the LF normalization 
($\Delta\phistar\approx0.15$) and
characteristic magnitude ($\Delta\Mstar\approx0.22$).  The combined
\UDFDeep~and~the \GOODSERS~survey also produce relatively low correlation
coefficients ($\rho\approx0.6-0.85$) compared with other surveys 
combinations; the orientation of the $\Mstar-\alpha$ joint constraint
from \UDFDeep~is only slightly inclined ($\oangle=12\deg$), and allows
the \GOODSERS~survey ($\oangle=55\deg$) to improve the combined 
constraints on $\Mstar$.

\subsection{\GOODSAll}

These model surveys illustrate how increasing the survey limiting depth
over a moderate area alter the forecasted LF parameter constraints.
These surveys share a common total area ($\Atot=326$ square arcmin) and
number of fields ($\Nfields=2$), but span a factor of two in integration
time ($\approx\sqrt{2}$ in sensitivity, $\HAB=28.2-28.6$, see Table 
\ref{table:model_surveys}).  Because the extra depth probes more abundant,
lower-luminosity galaxies, the typical galaxy bias 
($\bave\approx6.5$) and cosmic variance uncertainty ($\sigmaCV\approx0.142$)
in \GOODSDeep~would be smaller 
than for either \GOODSMedium~($\bave\approx6.7$, $\sigmaCV\approx0.146$)
or \GOODSShallow~($\bave\approx7.0$, $\sigmaCV\approx0.152$).
Similarly, the extra depth affords more observed galaxies ($N\approx1100$) and
less Poisson uncertainty ($3\%$ fractionally) for \GOODSDeep~than for
\GOODSMedium~($N\approx850$, $3.4\%$) or \GOODSShallow~($N\approx610$, $4.1\%$).

The Fisher matrix calculations translate the Poisson and cosmic variance
uncertainties into constraints on the LF parameters, and 
Figures \ref{fig:survey_udf_go_model_goods_deep}-\ref{fig:survey_udf_go_model_goods_shallow}
show how the parameter constraints scale with limiting magnitude for
\GOODSAll.
In each figure, the
shaded areas show the constraints achieved by \UDFGO~(light blue),
the model surveys (blue), and the combination of \UDFGO~and each
model survey (dark blue).  The LF parameter constraints are also listed
in Table \ref{table:model_survey_constraints}.

Of these three surveys, \GOODSDeep~achieves the best combined parameters constraints 
($\Delta\Mstar=0.21$, $\Delta\logten\phistar=0.176$, $\Delta\alpha=0.172$).  However,
the relative gain over \GOODSMedium~($\Delta\Mstar=0.24$, $\Delta\logten\phistar=0.19$, 
$\Delta\alpha=0.21$) and \GOODSShallow~($\Delta\Mstar=0.25$, $\Delta\logten\phistar=0.21$, 
$\Delta\alpha=0.25$) are relatively modest ($20\%$ improvement in $\Delta\Mstar$ and 
$\Delta\logten\phistar$, and $40\%$ in $\Delta\alpha$).  Most of the relative improvement
owes to the increased constraint on the LF faint-end slope for \GOODSDeep, since
the three surveys are essentially identical for galaxies with $\HAB<28.2$, have
a similar orientation of their error ellipse in the $\Mstar$-$\phistar$ projections
($\oangle\approx39\deg$), and have similar correlations between LF parameters.
Combining with \UDFGO~results in a larger relative improvement in the LF parameter
constraints for \GOODSMedium~(10\%) and \GOODSShallow~(20-25\%) than for \GOODSDeep~(5\%).

\subsection{\SEDS}

The next widest model survey design is \SEDS, with a total area of 
$\Atot=967.6$ square arcmin to $\HAB=27.8$ depth over $\Nfields=4$ fields.  
Such a survey
would find $N\approx940$ galaxies at $z\sim7$, with an average
bias of $\bave=7.5$, cosmic variance uncertainty of $\sigmaCV\approx0.1$,
and Poisson uncertainty of 3\%.

Figure \ref{fig:survey_udf_go_goods_ers_model_seds} shows the constraints for
\SEDS~(blue region), the combination of \UDFGO~and \GOODSERS~(light blue region), 
and the combination of all three surveys (dark blue region).
The larger area of \SEDS~allows for better or comparable combined constraints on the LF characteristic
magnitude ($\Delta\Mstar\approx0.19$) and normalization ($\Delta\logten\phistar\approx0.15$)
than deeper surveys over smaller areas.  Owing to its weaker constraint on 
the faint-end slope, the error ellipses
provided by \SEDS~are more highly inclined in the $\Mstar$-$\alpha$ ($\oangle\approx50\deg$) and
$\phistar$-$\alpha$ ($\oangle\approx59\deg$) projections than the \UDFGO-\GOODSERS~combined
constraints ($38$ and $44\deg$).
When combined with \UDFGO~and \GOODSERS~surveys, \SEDS~would provide among
the tightest constraints of the surveys we consider (with \UDFDeep~providing better
combined constraints on $\alpha$ and \COSMOS~providing better constraints on $\Mstar$,
$\phistar$, and $\alpha$).

Of additional interest for a design like \SEDS~is some measure of the benefit of having $\Nfields=4$
for constraining the $z\sim7$ LF compared to a single contiguous field.  We note
that changing \SEDS~to a single field of the same total area and aspect ratio results
in essentially no change to the constraints on the LF parameters (a fractional change of
less than 1\%).  The cosmic
variance uncertainty does improve by $\sim25\%$ (see Figure \ref{fig:multiple_fields}) from
$\sigmaCV\approx0.13$ when increasing the number of fields from $\Nfields=1$ to $\Nfields=4$,
but this improvement has little net effect on the LF parameter constraints.
The marginalized constraints on the LF parameters are sensitive to the Poisson
errors of individual magnitude bins on the bright end of the LF, and the Poisson
error is independent of $\Nfields$ for surveys of fixed total area.  For magnitude bins that are
Poisson-uncertainty dominated, the improvement in the cosmic sample variance gained
by increasing $\Nfields$ therefore may not strongly influence end constraints on the LF parameters.

\subsection{\COSMOS}

\begin{figure*}
\figurenum{9}
\epsscale{1.1}
\plotone{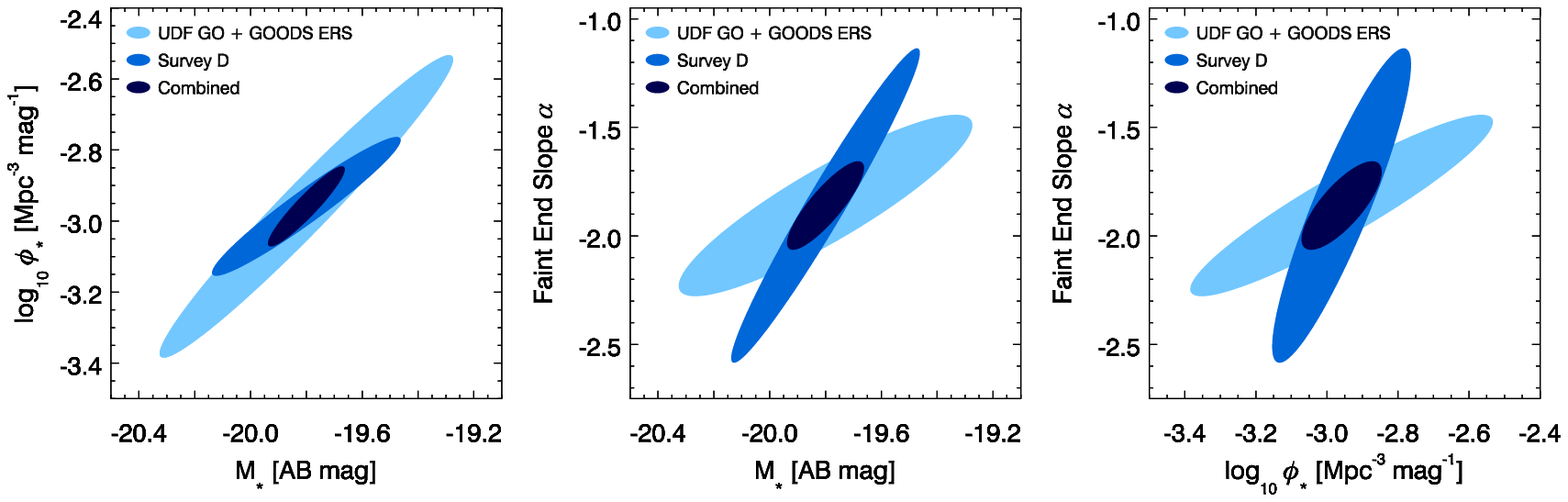}
\caption{\label{fig:survey_udf_go_goods_ers_model_cosmos}
Forecasted constraints on $z\sim7$ luminosity function parameters expected from the existing \UDFGO~and \GOODSERS~surveys and the model \COSMOS.
The constraints are calculated for a \citet{schechter1976a} luminosity function with a characteristic luminosity $\Mstar$, normalization
$\phistar$, and faint-end slope $\alpha$.
Shown are the $1-\sigma$ constraints in the $\Mstar-\logten \phistar$ (left panel), $\Mstar-\alpha$ (middle panel), and $\logten \phistar-\alpha$ (right 
panel) space projections for the combined \UDFGO~and \GOODSERS~surveys (light blue region) and the model \COSMOS~(blue region).  Also 
shown are the constraints
expected by combining all three surveys (dark blue region).  
}
\end{figure*}

The widest and shallowest model survey design considered is the single-field
\COSMOS~($\Atot\approx1\deg^{2}$, $\HAB=27$, $\Nfields=1$).  This model survey
would find $N\approx570$ galaxies at $z\sim7$, probing only galaxies brighter
than $\Mstar$ with an average bias of $\bave\approx9$ with a cosmic variance
uncertainty of $\sigmaCV\approx0.11$ (dominating over the Poisson uncertainty
of $4.2\%$).  Figure \ref{fig:survey_udf_go_goods_ers_model_cosmos} shows
the constraints that would be achieved by the combination of \UDFGO~and \GOODSERS~
(light blue region), \COSMOS~individually (blue region), and the combination
of all three surveys (dark blue region).

The constraints achievable by \COSMOS~individually are comparable to the constraints
provided by combining \UDFGO~and \GOODSERS, but would require roughly three times
as much additional telescope time to complete.  However, the combination of 
\COSMOS~with both \UDFGO~and \GOODSERS~produces the strongest joint constraint of any
survey design we considered ($\Delta\Mstar\approx0.136$, 
$\Delta\logten\phistar\approx0.11$, $\Delta\alpha\approx0.20$).  
The orientation of constraint provided by \COSMOS~individually is inclined
($\boangle=[29.5, 65.4, 76.2]$) relative to the \UDFGO-\GOODSERS~combination
($\boangle=[38.9, 37.6, 44.3]$), and results in a relatively low correlation 
between the LF normalization and faint-end slope ($\rho\approx0.77$).
While other
survey designs produce better constraints on the faint-end slope, the joint constraint
region shown in Figure \ref{fig:survey_udf_go_goods_ers_model_cosmos} 
produces an uncertainty in the LF that
is better than $\approx6\%$ at all relatively bright ($\HAB\lesssim28$) magnitudes.

%%%%%%%%%%%%%%%%%%%%%%%%%%%%%%%%%%%%%%
%
%	Discussion
%
%%%%%%%%%%%%%%%%%%%%%%%%%%%%%%%%%%%%%%
\section{Discussion}
\label{section:discussion}

We have considered the problem of forecasting
constraints on parameters of the $z\sim7$ LF 
given the characteristics of on-going surveys
and models for potential future survey designs.
The purview of our calculation was purposefully
narrow since a more comprehensive evaluation of 
galaxy surveys could involve many 
additional questions we have not addressed.
We now turn to a variety of possible caveats
that stem from considering photometric galaxy 
survey designs
more generally.

We have focused on forecasting constraints
for the luminosity function.  Our approach
was modeled after Fisher matrix calculations
that used the abundance of galaxy clusters
to forecast cosmological parameters constraints
\citep{hu2003a,lima2004a,lima2005a,cunha2009a,wu2009a}, 
but other previous 
calculations have forecasted cosmological
parameter constraints from galaxy clustering
\citep[e.g.,][]{vogeley1996a,matsubara2001a,matsubara2003a,linder2003a,albrecht2009a}.
The incorporation of galaxy clustering data
can circumvent some assumptions made in 
\S \ref{section:variances} when using
simple abundance matching to assign galaxy
bias by replacing the sample variance
estimates in Equations \ref{eqn:sample_covariance}
and \ref{eqn:cosmic_variance} by an integral 
over the galaxy correlation function.
Other estimates of how cosmic variance
uncertainty is influenced by galaxy bias
have taken a similar approach 
\citep[e.g.,][]{newman2002a,somerville2004a,stark2007b,trenti2008a}.

Our calculations have regarded a limited
but interesting redshift regime near $z\sim7$.
While we have found that the combination of
existing deep/narrow surveys with a future wide/shallow
survey or a future ultradeep/narrow survey would
provide tight constraints on the $z\sim7$ luminosity
function,
studies of the 
galaxy population at higher and lower
redshifts could require substantially different
surveys.  For instance, the $z\gtrsim8$ dropout
candidates identified in the \UDFGO~data
are all fainter than $\HAB=27.7$  
\citep{bouwens2009a,bunker2009a,mclure2009a,yan2009a}.
The decreasing abundance of relatively bright galaxies
with increasing redshift will tend to favor deeper and
narrow surveys.   Our calculations can easily
be extended to estimate the constraining power of
various surveys designs for higher-redshift
galaxy populations, but we will save such estimates
for future work when better fiducial estimates of the
$z\gtrsim8$ luminosity function are available.

Our Fisher matrix approach requires
the use of a fiducial model for the abundance
of $z\sim7$.  We adopt the \cite{oesch2010a}
estimate of the galaxy luminosity function,
which was determined by scaling the
characteristic magnitude $\Mstar$ 
and normalization $\phistar$ from lower redshift data and 
then fitting for the faint-end slope $\alpha$.
If the $z\sim7$ galaxy luminosity function
differs substantially from the \cite{oesch2010a}
estimate, then our forecasted constraints could
be similarly inaccurate.  For instance, if
the normalization $\phistar$ was considerably
lower or the characteristic
magnitude $\Mstar$ much fainter than
that estimate by \cite{oesch2010a}, then the
relative benefit of combining
the \UDFGO~and \GOODSERS~data
with a wide/shallow survey over a narrow/ultradeep 
design could be reduced.  

The calculations in \S \ref{subsubsection:multiple_fields}
and 
\S \ref{section:results} suggest that
splitting wide surveys into multiple fields
to probe statistically-independent regions 
of the universe may not 
dramatically improve constraints on the
galaxy luminosity function.  While this
conclusion depends strongly on the total 
volume of the survey, other considerations
such as scheduling, field observability,
or sky backgrounds could make multiple
fields advantageous compared with a 
single contiguous field of the same total
area.  

The abundance matching calculation also
requires either knowledge or assumption 
about the completeness of the survey and
the fraction of dark matter halos occupied
by galaxies.  We have assumed that
the surveys are essentially volume-limited
and that there is a one-to-one correspondence
between galaxies and dark matter halos.
Both of these assumptions are likely imperfect,
and some estimates of the high-redshift occupation fraction
are as low as $20\%$ \citep{stark2007b}.
The influence of these assumptions over the 
forecasted parameter
constraints depends on the character of the survey.
For a given observed luminosity function, reducing 
the halo occupation fraction or the survey
completeness acts to reduce the effective galaxy bias
in the survey by either increasing the number of
halos per galaxy in the survey or increasing the
number of undetected galaxies.  In either case, the
Poisson uncertainty is based on the observed number
of galaxies and is unaffected.  For purposes of
constraining the observed luminosity function,
wide surveys are fairly insensitive to either assumption,
since Poisson uncertainty in the abundance of bright
galaxies plays a large role in their error budget.
The change in sample variance for narrow surveys 
can lead to a degradation of the marginalized parameter
constraints (while the unmarginalized constraints can improve) 
by increasing the parameter correlations.  However, the
relative effect is small and the degradation is only
$2\times$ for simultaneously low 
incompleteness ($\fcomp=0.1$) and 
small occupation fraction ($\focc=0.1$).  

%%%%%%%%%%%%%%%%%%%%%%%%%%%%%%%%%%%%%%
%
%	Summary
%
%%%%%%%%%%%%%%%%%%%%%%%%%%%%%%%%%%%%%%
\section{Summary}
\label{section:summary}

Motivated by the exciting initial galaxy
survey data obtained
by newly-installed Wide Field Camera 3 (WFC3)
on the Hubble Space Telescope (HST), we have
attempted to quantify how well on-going and
possible future infrared surveys with WFC3 will constrain
the abundance of galaxies at $z\sim7$.
Our primary methods and results include:

\begin{itemize}

\item We perform a  
Fisher matrix calculation to forecast constraints
on the galaxy luminosity function (LF) achievable
by a survey with a given depth, area, and
number of fields.  In our approach, the 
constraints on the
LF normalization $\phistar$,
characteristic magnitude $\Mstar$, and
faint-end slope $\alpha$ that a survey can
achieve directly relate to the sample cosmic
variance and Poisson uncertainty
on the observed galaxy abundance through
the Fisher matrix.  For a fiducial LF
model, the abundance of observed
galaxies and dark matter halos are matched
\citep[e.g.,][]{conroy2006a,conroy2009a}
to estimate the bias of galaxies of a given
luminosity.  The galaxy bias is combined
with the RMS density fluctuations within
the survey volume to calculate the sample cosmic
variance \citep[e.g.,][]{newman2002a,somerville2004a,stark2007b,trenti2008a}, 
while the Poisson variance simply
scales with the square root of the number of
of observed galaxies.  The constraining
power of each survey is then calculated from
the Fisher matrix using the \cite{schechter1976a} 
model of the LF, its derivatives,
the sample cosmic variance and Poisson uncertainties,
and any data covariance.
The combined constraints from multiple surveys can
be estimated easily by summing their Fisher matrices.
Similar calculations should prove useful for designing
future photometric surveys and estimating their constraining
power for the galaxy LF.

\item Using the Fisher matrix calculations,
we estimate the constraints on the abundance
of $z\sim7$ galaxies that will be
achieved with the entire forthcoming Ultradeep
Field Guest Observation (\UDFGO)
and Great Observatories Origins Deep Survey Early-Release
Science (\GOODSERS) HST WFC3 IR channel data.  
Using the $z\sim7$ galaxy LF estimated by \cite{oesch2010a}
as a fiducial model, we calculate that the 
combined \UDFGO~and \GOODSERS~data will 
achieve
marginalized (unmarginalized) LF parameter constraints of
$\Delta\Mstar\approx0.5$ mag (0.1 mag), $\Delta\logten\phistar\approx0.4$
(0.1), and $\Delta\alpha\approx0.4$ (0.15) when
the surveys are fully completed.  These marginalized
constraints correspond to uncertainties in the
total number of $z\sim7$ galaxies with magnitudes
$\MUV<-18$ ($\MUV<\Mstar\approx-19.8$) of 25\% (200\%),
after accounting for covariances between the LF parameters.
These surveys will provide the first detailed information
on $z\sim7$ galaxy populations, but abundance of the
bright-end of the $z\sim7$ LF will remain uncertain without
further data.

\item We also forecast $z\sim7$ LF constraints provided
by a variety of model WFC3 surveys that
would each require $\sim450-900$ HST orbits.  The six
model surveys considered cover a large range of areas ($14-3600$ square
arcmin) and depths ($\HAB=27-30$ in $\FFOURTY$) to
study the relative value area and depth for constraining
the abundance of $z\sim7$ galaxies.  When combined with
the forthcoming \UDFGO~and \GOODSERS~data, all the surveys
considered produce interesting luminosity function constraints
(see Table \ref{table:model_survey_constraints}).
We find that a $\sim1\deg^{2}$
survey to $\HAB\approx27$ in \FFOURTY~provides the tightest
combined marginalized constraints ($\Delta\Mstar\approx0.14$, 
$\Delta\phistar\approx0.11$, $\Delta\alpha\approx0.20$) on
the abundance of $z\sim7$ galaxies of all survey designs
we consider, but only by a small margin.  This survey 
would require 780 total orbits, including
equal coverage in a bluer WFC3 filter to define a drop out color
selection.  In contrast, the abundance of \emph{faint} galaxies would be best
constrained by increasing depth of the HST ultradeep fields to 
$\sim125$ orbits per pointing ($\HAB\approx30.1$ in \FFOURTY), which
provides marginalized LF constraints of 
$\Delta\Mstar\approx0.22$, 
$\Delta\phistar\approx0.15$, and $\Delta\alpha\approx0.10$ for
750 total orbits (including equal coverage in a bluer WFC3 filter).

\item We also consider the usefulness of splitting surveys into
$\Nfields$ multiple fields to probe independent samples and reduce cosmic
variance uncertainties \citep[e.g.,][]{newman2002a}.  We show that
the shape of the \LCDM power spectrum limits the statistical gain
of splitting a high-redshift survey into multiple fields to $\lesssim10\%$ 
(for $\Nfields=2$)
when the survey area is large ($\gtrsim0.5\deg^{2}$).  We suggest
that this statistical gain should be weighed against any scientific
gains achieved by probing large contiguous areas.

\end{itemize}

Initial analyses of the \UDFGO~data have already demonstrated
that the installation of WFC3 on HST will transform our knowledge
of high-redshift galaxy populations at $z\sim7$  and beyond
\citep[e.g.,][]{bouwens2009a,bouwens2010b,bouwens2010a,oesch2010a,oesch2010b,bunker2009a,mclure2009a,yan2009a,wilkins2009a,labbe2009a,labbe2010a,finkelstein2009a}.
Our work has attempted to quantify expectations for the constraining
power of the \UDFGO~and \GOODSERS~surveys, and forecast constraints
achievable with more extensive future surveys using WFC3 or other
instruments.
These calculations illustrate how truly powerful the refurbished HST
is for exploring high-redshift galaxy populations, and emphasize
how exciting near-term gains in our knowledge of $z\gtrsim7$ galaxies 
will be.

\acknowledgements
I thank Peter Capak and Nick Scoville for useful advice 
on modeling the survey data and WFC3 observations, 
Richard Ellis for helpful discussions, and
Anatoly Klypin for permission to use his cosmological 
simulation.  I am
supported by a Hubble Fellowship grant, program number 
HST-HF-51262.01-A provided by NASA from the Space Telescope 
Science Institute, which is operated by the Association of 
Universities for Research in Astronomy, Incorporated, 
under NASA contract NAS5-26555.  I also thank the Caltech
Astronomy Department for hosting me during my Hubble
Fellowship.
\\

\clearpage

\clearpage

\begin{deluxetable*}{lcccccc}
\tablecolumns{7} 
\tablewidth{0pc} 
\tablecaption{Forecasted Constraints for Existing Surveys} 
\tablehead{
\colhead{Survey} & \colhead{Marg. Uncert.} & \colhead{Unmarg. Uncert.} & \colhead{Marg. Uncert.}       & \colhead{Unmarg. Uncert.}       & \colhead{Marg. Uncert.} & \colhead{Unmarg. Uncert.}\\
\colhead{Name}   & \colhead{$\Mstar$}      & \colhead{$\Mstar$}        & \colhead{$\logten \phistar$}  & \colhead{$\logten \phistar$}    & \colhead{$\alpha$}      & \colhead{$\alpha$}\\
\label{table:existing_survey_constraints}}
\startdata 
\UDFGO    &0.837 &  0.160 &  0.655 &  0.093 &  0.566 &  0.162\\
\GOODSERS &1.338 &  0.171 &  1.026 &  0.145 &  1.852 &  0.520\\
Combined  &0.524 &  0.117 &  0.425 &  0.078 &  0.415 &  0.154\\
%\cutinhead{Survey & Pearson $\rho$ & Pearson $\rho$ & Pearson $\rho$ & $\oangle(\Mstar-\phistar)$ & $\oangle(\Mstar-\alpha)$ & $\oangle(\phistar-\alpha)$ }
%\cutinhead{
%Name   & $\Mstar$-$\logten \phistar$ & $\Mstar$-$\alpha$ & $\logten \phistar$ $\alpha$ & $[\deg]$ & $[\deg]$ & $[\deg]$ \\
\hline
\hline\\
\colhead{Survey} & \colhead{Pearson $\rho$}              & \colhead{Pearson $\rho$}    & \colhead{Pearson $\rho$}              & \colhead{$\oangle(\Mstar-\phistar)$} & \colhead{$\oangle(\Mstar-\alpha)$} & \colhead{$\oangle(\phistar-\alpha)$}\\
\colhead{Name}   & \colhead{$\Mstar$-$\logten \phistar$} & \colhead{$\Mstar$-$\alpha$} & \colhead{$\logten \phistar$-$\alpha$} & \colhead{$[\deg]$}                   & \colhead{$[\deg]$}                 & \colhead{$[\deg]$} \\ \\
\hline\\
\UDFGO    &0.98 &  0.91 &  0.95 &  38 &  33 &  41\\
\GOODSERS &0.99 &  0.96 &  0.95 &  37 &  55 &  62\\
Combined  &0.97 &  0.89 &  0.93 &  39 &  38 &  44
\enddata 
\end{deluxetable*}

%\begin{deluxetable*}{lcccccccccccc}
%\tablecolumns{13} 
%\tablewidth{0pc} 
%\tablecaption{Forecasted Constraints for Existing Surveys} 
%\tablehead{
%\colhead{Survey} & \colhead{Marg. Uncert.} & \colhead{Unmarg. Uncert.} & \colhead{Marg. Uncert.} & \colhead{Unmarg. Uncert.} & \colhead{Marg. Uncert.} & \colhead{Unmarg. Uncert.} & \colhead{Pearson $\rho$}       & \colhead{Pearson $\rho$}     & \colhead{Pearson $\rho$} & \colhead{$\oangle(\Mstar-\phistar)$} & \colhead{$\oangle(\Mstar-\alpha)$} & \colhead{$\oangle(\phistar-\alpha)$}\\
%\colhead{Name}   & \colhead{$\Mstar$}     & \colhead{$\Mstar$}     & \colhead{$\logten \phistar$}  & \colhead{$\logten \phistar$}    & \colhead{$\alpha$}    & \colhead{$\alpha$}    & \colhead{$\Mstar$-$\logten \phistar$} & \colhead{$\Mstar$-$\alpha$} & \colhead{$\logten \phistar$-$\alpha$} & \colhead{$[\deg]$} & \colhead{$[\deg]$} & \colhead{$[\deg]$} 
%\label{table:existing_survey_constraints}}
%\startdata 
%\UDFGO    &0.837 &  0.160 &  0.655 &  0.093 &  0.566 &  0.162 &  0.98 &  0.91 &  0.95 &  38 &  33 &  41\\
%\GOODSERS &1.338 &  0.171 &  1.026 &  0.145 &  1.852 &  0.520 &  0.99 &  0.96 &  0.95 &  37 &  55 &  62\\
%Combined  &0.524 &  0.117 &  0.425 &  0.078 &  0.415 &  0.154 &  0.97 &  0.89 &  0.93 &  39 &  38 &  44
%\enddata 
%\end{deluxetable*}

\begin{deluxetable*}{lcccccc}
\tablecolumns{7} 
\tablewidth{0pc} 
\tablecaption{Forecasted Constraints for Model Surveys} 
\tablehead{
\colhead{Survey} & \colhead{Marg. Uncert.} & \colhead{Unmarg. Uncert.} & \colhead{Marg. Uncert.} & \colhead{Unmarg. Uncert.} & \colhead{Marg. Uncert.} & \colhead{Unmarg. Uncert.}\\
\colhead{Name}   & \colhead{$\Mstar$}     & \colhead{$\Mstar$}     & \colhead{$\logten \phistar$}  & \colhead{$\logten \phistar$}    & \colhead{$\alpha$}    & \colhead{$\alpha$}
\label{table:model_survey_constraints}}
\startdata 
\UDFDeep& 0.276 &  0.158 &  0.163 &  0.064 &  0.104 &  0.057\\
\GOODSDeep& 0.225 &  0.084 &  0.185 &  0.053 &  0.182 &  0.055\\
\GOODSMedium& 0.257 &  0.084 &  0.210 &  0.057 &  0.231 &  0.068\\
\GOODSShallow& 0.299 &  0.085 &  0.244 &  0.061 &  0.305 &  0.089\\
\SEDS& 0.244 &  0.052 &  0.193 &  0.041 &  0.311 &  0.088\\
\COSMOS& 0.336 &  0.041 &  0.194 &  0.045 &  0.722 &  0.137\\
Ex.+\UDFDeep\tablenotemark{a}& 0.221 &  0.116 &  0.151 &  0.058 &  0.101 &  0.057\\
Ex.+\GOODSDeep\tablenotemark{b}& 0.214 &  0.074 &  0.176 &  0.046 &  0.172 &  0.052\\
Ex.+\GOODSMedium\tablenotemark{b}& 0.236 &  0.074 &  0.194 &  0.049 &  0.207 &  0.063\\
Ex.+\GOODSShallow\tablenotemark{b}& 0.253 &  0.075 &  0.210 &  0.051 &  0.245 &  0.078\\
Ex.+\SEDS\tablenotemark{c}& 0.185 &  0.048 &  0.150 &  0.036 &  0.219 &  0.077\\
Ex.+\COSMOS\tablenotemark{c}& 0.136 &  0.039 &  0.111 &  0.039 &  0.201 &  0.103\\
\hline
\hline\\
\colhead{Survey} & \colhead{Pearson $\rho$}       & \colhead{Pearson $\rho$}     & \colhead{Pearson $\rho$} & \colhead{$\oangle(\Mstar-\phistar)$} & \colhead{$\oangle(\Mstar-\alpha)$} & \colhead{$\oangle(\phistar-\alpha)$}\\
\colhead{Name}   & \colhead{$\Mstar$-$\logten \phistar$} & \colhead{$\Mstar$-$\alpha$} & \colhead{$\logten \phistar$-$\alpha$} & \colhead{$[\deg]$} & \colhead{$[\deg]$} & \colhead{$[\deg]$} \\ \\
\hline\\
\UDFDeep&0.79 &  0.52 &  0.81 &  28 &  12 &  30\\
\GOODSDeep&0.92 &  0.91 &  0.95 &  39 &  38 &  45\\
\GOODSMedium&0.94 &  0.93 &  0.95 &  39 &  42 &  48\\
\GOODSShallow&0.95 &  0.94 &  0.95 &  39 &  46 &  52\\
\SEDS&0.97 &  0.95 &  0.95 &  38 &  52 &  59\\
\COSMOS&0.96 &  0.97 &  0.90 &  29 &  65 &  76\\
Ex.+\UDFDeep\tablenotemark{a}& 0.84 &  0.62 &  0.82 &  33 &  18 &  31\\
Ex.+\GOODSDeep\tablenotemark{b}&0.94 &  0.91 &  0.95 &  39 &  38 &  44\\
Ex.+\GOODSMedium\tablenotemark{b}&0.95 &  0.92 &  0.95 &  39 &  41 &  47\\
Ex.+\GOODSShallow\tablenotemark{b}&0.95 &  0.92 &  0.95 &  39 &  44 &  50\\
Ex.+\SEDS\tablenotemark{c}&0.96 &  0.92 &  0.93 &  39 &  50 &  56\\
Ex.+\COSMOS\tablenotemark{c}&0.93 &  0.86 &  0.77 &  39 &  58 &  65
\enddata 
\tablenotetext{a}{Combined with the existing \GOODSERS~survey.}
\tablenotetext{b}{Combined with the existing \UDFGO~survey.}
\tablenotetext{c}{Combined with the existing \UDFGO~and \GOODSERS~surveys.}
\end{deluxetable*}

\end{document}